\documentclass[aps,groupedaddress,floatfix,nofootinbib,preprintnumbers,eqsecnum,twocolumn]{revtex4}
\usepackage{amssymb,amsmath,graphicx,multirow,epsf}
\usepackage{dcolumn}   
\usepackage{bm}        

\begin{document}


\title{Overcoming Velocity Suppression in Dark-Matter Direct-Detection Experiments\\}
\author{Keith R. Dienes$^{1,2}$\footnote{E-mail address:  {\tt dienes@email.arizona.edu}},
      Jason Kumar$^{3}$\footnote{E-mail address:  {\tt jkumar@hawaii.edu}},
      Brooks Thomas$^{4}$\footnote{E-mail address:  {\tt bthomas@physics.carleton.ca}},
      David Yaylali$^{3}$\footnote{E-mail address:  {\tt yaylali@hawaii.edu}}}
\affiliation{
     $^1\,$Department of Physics, University of Arizona, Tucson, AZ  85721  USA\\
     $^2\,$Department of Physics, University of Maryland, College Park, MD  20742  USA\\
     $^3\,$Department of Physics, University of Hawaii, Honolulu, HI 96822  USA\\
     $^4\,$Department of Physics, Carleton University, Ottawa, ON K1S 5B6 Canada}

\begin{abstract}
Pseudoscalar couplings between Standard-Model quarks and dark matter 
are normally not considered relevant for dark-matter direct-detection experiments
because they lead to velocity-suppressed scattering cross-sections in the non-relativistic limit.
However, at the nucleon level, such couplings are effectively enhanced 
by factors of order ${\cal O}(m_N/m_q)\sim 10^3$, where $m_N$ and $m_q$
are appropriate nucleon and quark masses respectively.   This enhancement
can thus be sufficient to overcome the corresponding velocity suppression,
implying --- contrary to common lore --- that direct-detection experiments can 
indeed be sensitive to pseudoscalar couplings.  In this work, we explain how this 
enhancement arises, and present a model-independent analysis of pseudoscalar 
interactions at direct-detection experiments.  We also 
identify those portions of 
the corresponding dark-matter parameter space
which can be probed at current and future experiments of this type, 
and discuss the role of isospin violation in enhancing 
the corresponding experimental reach.
\end{abstract}

\maketitle

\newcommand{\newc}{\newcommand}
\newc{\gsim}{\lower.7ex\hbox{$\;\stackrel{\textstyle>}{\sim}\;$}}
\newc{\lsim}{\lower.7ex\hbox{$\;\stackrel{\textstyle<}{\sim}\;$}}
\makeatletter
\newcommand{\biggg}{\bBigg@{3}}
\newcommand{\Biggg}{\bBigg@{4}}
\makeatother

\def\vac#1{{\bf \{{#1}\}}}

\def\beq{\begin{equation}}
\def\eeq{\end{equation}}
\def\beqn{\begin{eqnarray}}
\def\eeqn{\end{eqnarray}}
\def\calM{{\cal M}}
\def\calV{{\cal V}}
\def\calF{{\cal F}}
\def\half{{\textstyle{1\over 2}}}
\def\quarter{{\textstyle{1\over 4}}}
\def\ie{{\it i.e.}\/}
\def\eg{{\it e.g.}\/}
\def\etc{{\it etc}.\/}
\def\qbar{{ \overline{q} }}
\def\Nbar{{ \overline{N} }}
\def\chibar{{ \overline{\chi} }}
\def\psibar{{ \overline{\psi} }}


\def\inbar{\,\vrule height1.5ex width.4pt depth0pt}
\def\IR{\relax{\rm I\kern-.18em R}}
 \font\cmss=cmss10 \font\cmsss=cmss10 at 7pt
\def\IQ{\relax{\rm I\kern-.18em Q}}
\def\IZ{\relax\ifmmode\mathchoice
 {\hbox{\cmss Z\kern-.4em Z}}{\hbox{\cmss Z\kern-.4em Z}}
 {\lower.9pt\hbox{\cmsss Z\kern-.4em Z}}
 {\lower1.2pt\hbox{\cmsss Z\kern-.4em Z}}\else{\cmss Z\kern-.4em Z}\fi}
\def\OmegaDM{\Omega_{\mathrm{CDM}}}
\def\Omegatot{\Omega_{\mathrm{tot}}}
\def\rhocrit{\rho_{\mathrm{crit}}}
\def\arcsinh{\mbox{arcsinh}}
\def\BRgamma{\mathrm{BR}_{\lambda}^{(2\gamma)}}
\def\OmegaDM{\Omega_{\mathrm{CDM}}}
\def\tnow{t_{\mathrm{now}}}
\def\Omegatotnow{\Omega_{\mathrm{tot}}^\ast}
\def\erf{\mathrm{erf}}
\def\rhototloc{\rho^{\mathrm{loc}}_{\mathrm{tot}}}
\def\Ecut{E_{\mathrm{cut}}}
\def\Emax{E_{\mathrm{max}}}

\newcommand{\ipb}{\text{pb}^{-1}}
\newcommand{\ifb}{\text{fb}^{-1}}
\newcommand{\iab}{\text{ab}^{-1}}
\newcommand{\ev}{\text{eV}}
\newcommand{\kev}{\text{keV}}
\newcommand{\mev}{\text{MeV}}
\newcommand{\gev}{\text{GeV}}
\newcommand{\tev}{\text{TeV}}
\newcommand{\pb}{\text{pb}}
\newcommand{\mb}{\text{mb}}
\newcommand{\cm}{\text{cm}}
\newcommand{\m}{\text{m}}
\newcommand{\km}{\text{km}}
\newcommand{\kg}{\text{kg}}
\newcommand{\g}{\text{g}}
\newcommand{\s}{\text{s}}
\newcommand{\yr}{\text{yr}}
\newcommand{\Mpc}{\text{Mpc}}
\newcommand{\etal}{{\em et al.}}
\newcommand{\ibid}{{\em ibid.}}

\newcommand{\be}{\begin{equation}}
\newcommand{\ee}{\end{equation}}
\newcommand{\ba}{\begin{align}}
\newcommand{\ea}{\end{align}}

\hyphenation{ALPGEN}
\hyphenation{EVTGEN}
\hyphenation{PYTHIA}

\def\ie{{\it i.e.}\/}
\def\eg{{\it e.g.}\/}
\def\etc{{\it etc}.\/}
\def\Yeq{Y^{\mathrm{eq}}}
\def\peff{p_{\mathrm{eff}}}
\def\Weff{W_{\mathrm{eff}}}
\def\OmegaDM{\Omega_{\mathrm{DM}}}
\def\rhocrit{\rho_{\mathrm{crit}}}
\def\snow{s_{\mathrm{now}}}
\def\tnow{t_{\mathrm{now}}}
\def\wtM{\widetilde{M}}


\input epsf





\section{Introduction\label{sec:intro}}


Among the many different experimental approaches towards understanding 
the particle nature of dark matter, direct-detection experiments are 
the only ones which directly probe the actual scattering of dark matter against
ordinary Standard-Model (SM) matter.
In general, such experiments seek to observe the 
infrequent scatterings of
galactic-halo dark matter with atomic nuclei
by searching for unambiguous evidence
of the resulting nuclear recoils~\cite{GoodmanWitten:1985,JungmanKamionkowskiGriest,DirectDetReviews,LewinSmith}.
The discovery of such scattering events would arguably
provide the most compelling possible evidence for 
the existence of dark matter, and would represent a major step
towards the all-important goal of discerning
the nature of the underlying dynamics that connects dark matter to the visible world.

That direct-detection experiments are capable of such powers of discernment
is a direct consequence of the fact
that different coupling structures between dark matter and ordinary
Standard-Model matter   
yield significantly different scattering phenomenologies.
As a result, any analysis of the experimental prospects for a given
direct-detection experiment will inevitably rely on certain assumptions
concerning the types of couplings that lead to this scattering.
Since the interactions between the dark and visible sectors are by definition suppressed,
one well-motivated possibility is that these two sectors are coupled by high-scale dynamics
which gives rise to effective contact interactions at the energy scales relevant for
direct detection.
Although the consequences of couplings between dark matter and Standard-Model leptons 
have certainly been studied in the prior literature (see, {\it e.g.}\/, Ref.~\cite{FoxPoppitz:2009}),
it is far more common to consider
elastic contact interactions between dark matter 
and SM quarks or gluons~\cite{JungmanKamionkowskiGriest,DirectDetReviews}.
This preference is ultimately motivated by the recognition that 
direct-detection experiments are designed to capitalize on
scattering between dark matter and atomic nuclei.
In this paper we shall focus on couplings to quarks.
Indeed, for a (Dirac) fermionic dark-matter particle $\chi$,
such contact interactions typically involve bilinear coupling structures of
the form
\begin{equation}
  (\overline{\chi} \, \Gamma \, \chi) \, (\overline{q} \, \Gamma' \, q)~,
\label{introeq}
\end{equation}
where $q$ denotes a Standard-Model quark and 
where $\Gamma$ and $\Gamma'$ represent different possible choices of Dirac gamma-matrix
combinations
$\lbrace {\bf 1}$, $i \gamma^5$, $\gamma^\mu$, $\gamma^\mu \gamma^5$, $\sigma^{\mu\nu}\rbrace$.
Different choices for $\Gamma$ and $\Gamma'$
correspond to different Lorentz and parity properties for the underlying interactions, 
and can thus lead to drastically different dark-matter phenomenologies (and therefore different predictions 
for associated event rates) at direct-detection experiments.  
For this reason, coupling structures which lead to attractive phenomenologies and
greater event rates tend to be studied ubiquitously in the dark-matter literature, while those leading
to suppressed event rates are typically neglected.

Unfortunately, by neglecting certain operators within this class, we are leaving many ``stones unturned" in the hunt
for dark matter.  In particular, it may turn out that certain coupling structures which are
seemingly suppressed (and thus are not considered) are actually enhanced by other factors.
Such enhancements could conceivably overcome the apparent suppressions associated with these operators, 
implying that the contributions from such operators are not negligible after all.

In this work, we show that this is indeed the case for {\it pseudoscalar}\/
coupling structures between dark matter and SM particles.  The standard lore is that 
such coupling structures lead to direct-detection event rates which are suppressed relative to those associated
with similar axial-vector coupling structures 
by factors of the $\chi$/nucleus relative velocity
$v\sim {\cal O}(10^{-3})$.
However, 
one of the main points of this paper is to emphasize that 
there is a corresponding mitigating factor that can potentially overcome this velocity suppression:
the process of transitioning from
a fundamental pseudoscalar {\it quark}\/ coupling to an effective pseudoscalar {\it nucleon}\/ coupling
introduces into the corresponding dark-matter scattering rate 
additional factors of order
${\cal O}(m_N/m_q)\sim 10^3$, where $m_{q,N}$ are the masses of the corresponding quarks and nucleons. 
Such enhancements, for example, are not present for axial-vector
interactions,
which are in some ways the  closest cousins to the pseudoscalar interactions.
In addition, we find that both axial-vector and pseudoscalar couplings 
are further enhanced in cases in which the dark-matter couplings are ultimately isospin-violating,
with these enhancements becoming particularly striking in the case of pseudoscalar interactions. 
Thus, contrary to popular lore,  we conclude that pseudoscalar couplings between dark matter and Standard-Model matter
can indeed be probed at dark-matter direct-detection experiments.

This paper is organized as follows.
First, we discuss the origins of the quark-to-nucleon enhancement factor that emerges
for pseudoscalar interactions, and provide a careful analysis
of the corresponding uncertainties that are inherent in such calculations.
We also demonstrate that the possibility of isospin-violating pseudoscalar
interactions only enhances these couplings further.
We then proceed to present a model-independent analysis of pseudoscalar 
interactions at direct-detection experiments.  In so doing, we 
also identify those portions of 
the corresponding dark-matter parameter space
which can be probed at current and future experiments of this type.


\section{From quarks to nucleons:  Velocity suppression and
nucleon enhancement for pseudoscalar couplings\label{sec:NucleonCouplings}}


We begin by discussing the matrix elements and couplings that describe the
contact interactions between fermionic dark matter and ordinary
Standard-Model matter.
This will also serve to introduce our notation and provide a point of comparison
between interactions involving different Lorentz and parity structures.
Ultimately, we shall focus on the cases of axial-vector 
and pseudoscalar interactions.
It turns out that these two cases are closely related, yet have different resulting
phenomenologies.

\subsection{General preliminaries:  Quark- and nucleon-level matrix
elements and pseudoscalar velocity suppression}

In general, we shall assume that our dark matter
is a Dirac fermion $\chi$ whose dominant couplings to
the visible sector
are to Standard-Model quarks
through dimension-six four-fermi contact interactions 
described by Lagrangian operators of the bilinear form
\beq
  {\cal O}_{\chi q}^{(XY)}~=~  { c_q^{(XY)} \over \Lambda^2} \, (\chibar \Gamma^{X}\chi)\, (\qbar \Gamma^{Y}\/ q)~.
\label{operator}
\eeq
Here $q=u,d,s,...$ specifies a particular species of quark,
$c_q$ is the corresponding $\chi/q$ coupling,
and $\Lambda$ corresponds to the mass scale of the new (presumably flavor-diagonal) 
physics which might generate such an effective interaction.
The $\Gamma^{X,Y}$ factors are appropriate combinations of Dirac gamma-matrices,
with the $X$ and $Y$ indices ranging over the values $\lbrace {\rm S,P,V,A,T}\rbrace$
corresponding  to $\Gamma^{\rm (S)}\equiv {\bf 1}$ (scalar interaction),
$\Gamma^{\rm (P)}\equiv  i\gamma^5$ (pseudoscalar),
$\Gamma^{\rm (V)}\equiv  \gamma^\mu $ (vector),
$\Gamma^{\rm (A)}\equiv  \gamma^\mu \gamma^5 $ (axial vector),
and
$\Gamma^{\rm (T)}\equiv \sigma^{\mu\nu}$ (tensor)
respectively.
The form in Eq.~(\ref{operator})
respects $U(1)_{\rm EM}$ and $SU(3)_{\rm color}$, as required,
although $SU(2)_{\rm weak}$ is broken. 
This is appropriate for energy and momentum scales below the electroweak scale.
The operator in Eq.~(\ref{operator}) is also Lorentz invariant provided that 
$X$ and $Y$ are both chosen from the set $\lbrace {\rm S,P}\rbrace$,
the set $\lbrace {\rm V,A}\rbrace$, or $\lbrace {\rm T}\rbrace$;  
note that in this last case, there are actually two ways in which the spacetime
 indices on each tensor can be contracted 
(either $\sigma_{\mu\nu}\sigma^{\mu\nu}$ or 
$\epsilon_{\mu\nu\lambda\rho} \sigma^{\mu\nu}\sigma^{\lambda\rho}$)
when forming the Lorentz-invariant operator.
In general, the operator ${\cal O}^{(XY)}_{\chi q}$ will be CP-even 
in all Lorentz-invariant cases except when $XY= {\rm SP}$, ${\rm PS}$, or ${\rm TT}$ with a contraction
through the $\epsilon$-tensor.
 
In direct-detection experiments, these operators induce scattering between
the dark-matter fermion $\chi$ and the individual nucleons $N$ of the detector substrate.
The tree-level matrix element describing this $\chi/N$ scattering 
is therefore given by
\begin{equation}
  {\cal M}^{(XY)}_{\chi N}  ~=~ \sum_{q}\,\frac{c_q^{(XY)}}{\Lambda ^2}
      \, \langle \chi_f | \chibar \Gamma^X \chi|\chi_i\rangle
      \, \langle N_f | \qbar \Gamma^Y q |N_i \rangle~
\label{matrixelement}
\end{equation}
where $N$ denotes the particular nucleon species in question (either
proton $p$ or neutron $n$).
Note that because the dark matter is a $U(1)_{\rm EM}$ singlet,
$N_i$ and $N_f$ are both of the same species $N$ and possibly differ
only in their momenta and/or spins as the result of the scattering.
The same will be assumed true for $\chi_i$ and $\chi_f$, even in cases 
such as those in Refs.~\cite{MultiComponentBlock,DDM}
in which the dark sector
has multiple components.

In general, the nucleonic matrix element
of the quark current $\qbar \Gamma^Y q$ cannot be evaluated analytically
within a nucleonic background defined by $N_i$ and $N_f$.
Indeed, to do so would require a complete understanding of the 
manner in which the quark degrees of freedom
are directly mapped onto those of the nucleon through the non-perturbative process
of confinement.
However, it is conventional to make the assumption 
that the nucleonic matrix element of the quark current is
proportional to that of the corresponding {\it nucleon}\/ 
current in the limit of vanishing momentum transfer~\cite{Shifman,PittelVogel,JaffeManohar,EllisOliveSavage:2008}:
\begin{equation}
  \langle N_f | \overline{q}\Gamma^Y  q | N_i\rangle ~\equiv~ 
    \Delta q^{(N)}\,
  \langle N_f | \overline{N}\Gamma^Y  N | N_i\rangle~,
\label{eq:SpinFraction}
\end{equation}
where $\Delta q^{(N)}$ represents a fixed constant of proportionality that
encapsulates the non-perturbative physics inherent in low-energy QCD.~
Indeed, this constant of proportionality is assumed to depend on the 
quark and nucleon in question, and also the specific choice of
the Dirac-matrix structure $\Gamma^Y$ involved, but is otherwise assumed to be independent
of all other relevant variables (such as the particular spin and velocity
configurations of the initial and final $N_i$ and $N_f$ states).
In practice, the values of 
$\Delta q^{(N)}$ for the different relevant cases
are calculated numerically through lattice gauge-theory techniques
and/or extracted experimentally.
We should emphasize, however, that the relation in Eq.~(\ref{eq:SpinFraction}) holds only 
as an approximate phenomenological ``rule of thumb'', and comes with several correction terms
which can be taken to be small or even vanishing in various limits.  Further details can
be found in Ref.~\cite{Shifman}.
  
Given the numerical values of $\Delta q^{(N)}$ 
in Eq.~(\ref{eq:SpinFraction}), the rest of the matrix element
(\ref{matrixelement}) is now in a form which can be evaluated analytically.
We then find
\begin{equation}
  {\cal M}^{(XY)}_{\chi N}  ~=~ \frac{g_{\chi N}}{\Lambda ^2}
      \, \langle \chi_f | \chibar \Gamma^X \chi|\chi_i\rangle
      \, \langle N_f | \overline{N}\Gamma^Y  N | N_i\rangle~,
\label{matrixelement2}
\end{equation}
where the final dark-matter/nucleon coupling $g_{\chi N}$ is given by 
\beq
        g_{\chi N} ~\equiv~ \sum_q c_q ^{(XY)} \, \Delta q^{(N)}~.
\label{gchincouplings}
\eeq

In this paper, we shall be concerned with three particular Dirac-matrix
bilinears:  the scalar (S), the pseudoscalar (P), and the axial vector (A).
In the non-relativistic limit, the 
scalar bilinear matrix element behaves to leading order as
\beq
    {\rm S}\/:~~ \langle \psi_f | \psibar \psi | \psi_i\rangle ~\sim~
       2m_\psi\, (\xi_\psi^{s'})^\dagger \xi_\psi^{s}~,
\label{scalarnorm}
\eeq
where $\xi_\psi^s$ represents the two-component spinor corresponding to the fermion 
$\psi$ with spin $s$, and where $s$ and $s'$ represent the spins of $\psi_i$ and $\psi_f$
respectively.
By contrast, the corresponding pseudoscalar and axial-vector bilinear matrix elements
behave to leading order as
\beqn
    {\rm P}\/:&  & ~~\,\langle \psi_f | \psibar \gamma^5 \psi | \psi_i\rangle ~~~~~\sim~
       (\xi_\psi^{s'})^\dagger 
       \,[(\vec p_f -\vec p_i)\cdot \vec \sigma] \, \xi_\psi^{s}~ \nonumber\\
    {\rm A}\/:&  &  
      \begin{cases}
     \langle \psi_f | \psibar \gamma^0 \gamma^5 \psi | \psi_i\rangle &\sim~ 0\cr 
     \langle \psi_f | \psibar \vec\gamma \gamma^5 \psi | \psi_i\rangle &\sim~ 
       2m_\psi \, (\xi_\psi^{s'})^\dagger \vec\sigma \, \xi_\psi^{s}~,
        \end{cases}\nonumber\\
\label{limits}
\eeqn
where $\vec \sigma$ are the Pauli spin matrices.
Taking $\psi$ to correspond to our nucleon field $N$, we thus see that 
both the pseudoscalar and axial-vector 
cases lead to a {\it spin-dependent}\/ scattering amplitude to leading order.
It is for this reason that the coefficients $\Delta q^{(N)}$ for these cases
can be interpreted as characterizing the fraction of the spin 
of the nucleon $N$ that is carried by the quark $q$.
Indeed, in the case of pseudoscalar couplings, it is easy to show that
 {\it all}\/ terms --- and not just those at leading order --- are spin-dependent;
this follows directly from the symmetry-based observation that 
any CP-odd Lorentz-scalar quantity which 
depends on only the properties of the nucleon 
must involve the nucleon spin~\cite{ChackoSpinDep,JiJiMattLianTao,Freytsis:2010ne,Kumar:2013iva}.
On the other hand, 
we see that the pseudoscalar case also leads to a {\it velocity suppression}\/:
the corresponding 
matrix element in Eq.~(\ref{limits})
is proportional to the velocity
transfer $\Delta \vec v \equiv \vec v_f - \vec v_i$, which is ${\cal O}(10^{-3})$ for 
most regions of interest involving dark-matter particles originating in the galactic halo.
It is this velocity suppression which lies at the root of the
relative disregard for pseudoscalar interactions in 
the dark-matter literature.


\subsection{An enhancement factor for pseudoscalar matrix elements}
 

Given these observations, our next task is to determine the numerical values
of the 
$\Delta q^{(N)}$ coefficients for the different cases of interest.
In this paper, our interest in the scalar coupling structure will be restricted to
the dark-matter bilinear rather than the quark bilinear --- 
\ie, in the language of Eq.~(\ref{operator})
we will wish to consider the case
with $X={\rm S}\/$, but never $Y={\rm S}$.  Consequently,
we shall only require the values of the coefficients $\Delta q^{(N)}$ 
for the axial-vector ($Y={\rm A}$) and pseudoscalar ($Y={\rm P}$) cases.
We also emphasize that we are not merely interested in the ``central values'' of these coefficients;
we are also interested in understanding their associated statistical and experimental {\it uncertainties}\/.
As we shall see, it is only by keeping track of these uncertainties 
that we can make solid statements about the phenomenological consequences of the different
couplings in each case.

Historically, the numerical values of the $\Delta q^{(N)}$ coefficients for the axial-vector
case have been extracted through nucleon-structure scattering experiments~\cite{SMC,HERMES,COMPASS}
and through lattice gauge-theory calculations~\cite{QCDSF}.
The results that we shall use in this paper are quoted in Table~\ref{tab:Delta}, and represent
the most current values taken from experiment and theory. 
In this context, it is important to note that there are rather significant uncertainties 
associated with the values of the $\Delta q^{(N)}$.  While 
the measured values for $\Delta u^{(N)}$ and $\Delta d^{(N)}$ 
tend to agree reasonably well with results from 
lattice calculations, the values for $\Delta s^{(N)}$ obtained
using these two methods can differ quite significantly.  In this paper, we shall therefore 
adopt the  $\Delta u^{(N)}$ and $\Delta d^{(N)}$ values quoted in Ref.~\cite{QCDSF}, but choose 
values for the $\Delta s^{(N)}$ such that they lie between these lattice results and the 
experimentally measured values in Ref.~\cite{COMPASS}, roughly two standard deviations away 
from the central value obtained in each analysis. 

We also observe that the results quoted in Table~\ref{tab:Delta} respect
 {\it quark-level}\/ isospin invariance --- {\it i.e.}\/, they satisfy
\beq
      \Delta {u}^{(p,n)} = \Delta d^{(n,p)}~,~~~~
      \Delta {s}^{(p)} = \Delta s^{(n)}~.
\label{quarklevelisospin}
\eeq
This makes sense, as the results in Table~\ref{tab:Delta} are derived
in the limit in which the three light quarks are considered to be effectively massless.
Likewise, in this approximation, the remaining quarks are considered 
to be too heavy to contribute significantly to proton-level and neutron-level couplings.
Thus, in the axial-vector case, we shall additionally take
\beq
      \Delta c^{(p,n)} = 
      \Delta b^{(p,n)} = 
      \Delta t^{(p,n)} = 0~.
\eeq

\begin{table}
\begin{tabular}{||l | c  c||}
\hline
\hline
\multicolumn{2}{||c}{~~~~~~~~~~~~~~~~~$N=p$~} & ~~~~$N=n$~~~  \\ 
\hline
 ~~~$\Delta u^{(N)\phantom{^X}}$  &    $~~~\phantom{-}0.787 \pm 0.158$~~~ &        ~~~$-0.319 \pm 0.066$~~~ \\
 ~~~$\Delta d^{(N)}$~     &                 $-0.319 \pm 0.066$    & $\phantom{-}0.787 \pm 0.158$    \\
 ~~~$\Delta s^{(N)}$~     &          $-0.040 \pm 0.03\phantom{0}$ & $-0.040 \pm 0.03\phantom{0}$ \\
\hline
\hline
\end{tabular}
\caption{ ~Values used in this paper for the axial-vector coefficients $\Delta q^{(N)}$.
  The values for the $\Delta u^{(N)}$ and $\Delta d^{(N)}$ are taken from the recent lattice 
  results reported in Ref.~\cite{QCDSF}, while the values for the $\Delta s^{(N)}$ have been chosen
  such that they lie between these lattice results and the experimentally measured values in
  Ref.~\cite{COMPASS}, roughly two standard deviations away from the central value obtained 
  in each analysis.       
\label{tab:Delta}}
\end{table}

We now turn to consider the corresponding coefficients in the pseudoscalar case.
In order to distinguish these coefficients from the axial-vector coefficients above,
we shall denote the pseudoscalar coefficients as $\Delta \tilde q^{(N)}$.

Rather than representing an independent degree of freedom, 
it turns out~\cite{ChengPseudoscalar,BaiFoxHarnik:2010}
that the pseudoscalar coefficients $\Delta \tilde q^{(N)}$ can  
actually be determined theoretically in terms of the axial-vector coefficients $\Delta q^{(N)}$.
This is ultimately because a general axial-vector current 
$j^{\mu 5} \equiv \psibar \gamma^\mu \gamma^5 \psi$ is not conserved in a theory in
which $m_\psi\not =0$, but is instead related to 
the pseudoscalar current $j^5 \equiv \psibar i\gamma^5 \psi$ through a divergence relation of the form
\beq
      \partial_\mu  j^{\mu 5} ~=~ 2 m_\psi \,j^5  + {\alpha_s\over 4\pi} G_{\mu\nu}\tilde G^{\mu\nu}~,
\label{current}
\eeq
where the final term reflects the possible additional contribution to the  
non-conservation of $j^{\mu 5}$ coming from a chiral anomaly (such as the chiral anomaly of QCD).
Indeed, amongst all the fermion bilinears $\psibar \Gamma^Y \psi$ with which we started,
it is only the axial-vector and pseudoscalar bilinears which can be connected to each other
through such a direct relation.

It should be noted that in principle Eq.~(\ref{current}) also contains
additional contributions resulting from integrating out light hadron states such as the pion.
As discussed in Ref.~\cite{pionpole},
such a pion-induced additional contribution would appear as a pion pole term.
However, this contribution is relatively small because the relevant momentum transfers
for our analysis are in fact well below the pion mass.  Indeed, 
since we are studying spin-dependent scattering, we will be focusing on experiments 
(such as COUPP~\cite{COUPP4Limits}) which involve fluorine rather than xenon targets;
the corresponding momentum transfers are then smaller
because fluorine is lighter than xenon.
Moreover, it is often the case within such experiments that
events with large recoil energies are rejected due to the calibration difficulties
that exist in this regime~\cite{COUPPdiscussions}.
Thus, for all events of interest,
the resulting
momentum transfers are much smaller than in leading spin-independent direct-detection 
experiments,
and we may disregard such pion-induced pole terms in what follows.

Exploiting Eq.~(\ref{current}) and following Ref.~\cite{ChengPseudoscalar}, we can now proceed to derive 
an expression for the pseudoscalar coefficients $\Delta \tilde q^{(N)}$ 
in terms of the axial-vector coefficients $\Delta q^{(N)}$.
We begin by noting that
\beqn
  && m_N \Delta q^{(N)} \langle N_f | \Nbar i\gamma^5 N | N_i\rangle \nonumber\\
  && ~~~~~~~ =~ \half \Delta q^{(N)} \partial_\mu  \langle N_f | \Nbar \gamma^\mu \gamma^5 N | N_i\rangle \nonumber\\
  && ~~~~~~~ =~ \half \partial_\mu \left[ \Delta q^{(N)}  
                        \langle N_f | \Nbar \gamma^\mu \gamma^5 N | N_i\rangle \right]\nonumber\\
  && ~~~~~~~ =~ \half \partial_\mu \langle N_f | \qbar \gamma^\mu \gamma^5 q | N_i \rangle \nonumber\\
  && ~~~~~~~ =~ m_q \langle N_f | \qbar i\gamma^5 q | N_i \rangle 
            + {\alpha_s\over 8\pi} \langle N_f | G\tilde G| N_i \rangle~.~~~~~~~~~~~
\label{chain}
\eeqn
In Eq.~(\ref{chain}), the first equality follows from the current relation~(\ref{current})
in the nucleon-level theory, where 
(since all nucleons are color-neutral)
no QCD chiral anomaly exists. 
The second equality, by contrast, follows from the 
fact that the $\Delta q^{(N)}$ coefficients are presumed
to be constants without spacetime dependence,
while the third equality follows from the definition of $\Delta q^{(N)}$ as relating
the nucleon-level and quark-level axial-vector matrix elements.
The final equality then again follows from Eq.~(\ref{current}), 
now evaluated in the quark-level theory for which the QCD chiral anomaly is non-zero.

For each nucleon $N$, the relation in Eq.~(\ref{chain}) furnishes three constraint equations
(one for each of the light quarks $q=u,d,s$).  However, recognizing that our three desired 
coefficients $\Delta \tilde q^{(N)}$ are nothing but the ratios
between the $\langle N_f | \qbar i\gamma^5 q | N_i \rangle$ 
and $\langle N_f | \Nbar i\gamma^5 N | N_i\rangle$ 
matrix elements,
we see that we still have one unknown remaining, namely the matrix element involving the QCD anomaly.
An additional constraint equation is therefore called for.
Towards this end,
it is traditional (see, \eg, Ref.~\cite{ChengPseudoscalar}) to    
assume that the large-$N_c$ chiral limit is a valid approximation.
This then implies the additional constraint~\cite{Brodsky:1988ip}
\beq
    \langle N_f| \overline{u} \gamma^5 u | N_i \rangle
   + \langle N_f| \overline{d} \gamma^5 d | N_i \rangle
   + \langle N_f| \overline{s} \gamma^5 s | N_i \rangle ~=~ 0~.
\label{largeNc}
\eeq

In principle, we could then proceed with this as our remaining constraint equation.
However, the appeal to the large-$N_c$ limit introduces a rather significant new source
of uncertainties of order ${\cal O}(1/N_c)$ into our calculation.
Since we wish to keep track of these uncertainties in this paper,
we will ultimately need to find a way to parametrize the 
uncertainties inherent in the relation
(\ref{largeNc}).  
We shall therefore write Eq.~(\ref{largeNc}) in the modified form
\beqn
&&    \langle N_f| \overline{u} \gamma^5 u | N_i \rangle
   + \langle N_f| \overline{d} \gamma^5 d | N_i \rangle
   + \langle N_f| \overline{s} \gamma^5 s | N_i \rangle \nonumber\\
&& ~~~~~~~~~~~~~~~~~~~~~~=~ \eta \, \langle N_f| \Nbar \gamma^5 N | N_i\rangle~~~~~~~~~~~~~~
\label{largeNc2}
\eeqn
where the right side of this equation is designed to reflect this uncertainty, 
with the numerical coefficient $\eta$ assumed to have a vanishing central value  
but a relatively large uncertainty $\delta \eta \sim {\cal O}(1/N_c)$. 

This system of equations (\ref{chain}) and (\ref{largeNc2})
may now be solved for the coefficients $\Delta \tilde q^{(N)}
\equiv \langle N_f | \qbar i\gamma^5 q | N_i\rangle/ \langle N_f | \Nbar i\gamma^5 N | N_i\rangle$
as well as an analogous anomaly coefficient
\beq
\Delta \tilde G^{(N)} ~\equiv~ {\alpha_s\over 8\pi} \, 
 { \langle N_f | G\tilde G|N_i \rangle \over
  \langle N_f | \Nbar i\gamma^5 N |N_i \rangle}~. 
\eeq
The results are then given by
\begin{align}
  \Delta \tilde q^{(N)}& ~=~  \frac{m_N}{m_q}\left[\Delta q^{(N)} - X^{(N)}\right] \nonumber \\
  \Delta \tilde G^{(N)}& ~=~ m_N \, X^{(N)}~,
\label{eq:DeltaqPrimeEqs}
\end{align}
where we have defined
\begin{equation}
  X^{(N)} ~\equiv~ \Bigg(\sum_{q=u,d,s}\frac{1}{m_q}\Bigg)^{-1}
    \left[\Bigg(\sum_{q=u,d,s}\frac{\Delta q^{(N)}}{m_q}\Bigg) - \frac{\eta}{m_N} \right]~. 
\end{equation}
As we see in Eq.~(\ref{eq:DeltaqPrimeEqs}), the natural scale of the 
pseudoscalar $\Delta \tilde q^{(N)}$ coefficients
is greater than the natural scale of the
axial-vector 
$\Delta q^{(N)}$ coefficients
by a factor of $m_N/m_q$.
This effect thus tends to {\it enhance}\/ 
the pseudoscalar couplings relative to the axial-vector couplings,
thereby giving us hope that we might eventually be able to 
overcome the velocity suppression that afflicts the case of
pseudoscalar scattering.

It is perhaps worth pausing to discuss the theoretical origin of this enhancement factor.
In general, the definition of the $\Delta q^{(N)}$ coefficients in
Eq.~(\ref{eq:SpinFraction}) suggests that these coefficients are fractional quantities which describe
``how much'' of some physical quantity associated with the nucleon $N$ can be
attributed to a constituent quark $q$.  
For example, in the case of the axial-vector coefficients, this physical quantity
is spin, and the corresponding $\Delta q^{(N)}$ coefficient is known as a spin fraction.
 Na\"\i vely, this would lead one to expect that the quantities $\Delta q^{(N)}$ should
be relatively small, and certainly less than one.
However, there is also another feature whose effects are reflected 
in the magnitudes of these coefficients:  this is the difference in the intrinsic overall normalizations
associated with the quark and nucleon fields $q$ and $N$ respectively.
Indeed, as is conventional, each field $q$ or $N$ is normalized to its mass so that the corresponding state kets
will satisfy relations such as 
$\langle q | q \rangle=2m_q$ and 
$\langle N | N \rangle=2m_N$ [or equivalently, relations such as those in Eq.~(\ref{scalarnorm})].
Thus, quantities such as the $\Delta q^{(N)}$ 
coefficients which convert from quark currents to nucleon currents will also intrinsically
include factors that reflect this change in normalization.

Given this, it might be tempting to identify  
the pseudoscalar enhancement factor $m_N/m_q$ 
appearing in Eq.~(\ref{eq:DeltaqPrimeEqs}) 
as reflecting this second contribution, namely a change in normalization. 
However, we can easily see that this is {\it not}\/ the case:  the axial-vector
coefficients $\Delta q^{(N)}$ and the pseudoscalar coefficients $\Delta \tilde q^{(N)}$ 
each already intrinsically incorporate such normalization factors, yet our enhancement factor
in Eq.~(\ref{eq:DeltaqPrimeEqs}) is one which rescales our pseudoscalar coefficients
 {\it relative to the axial-vector coefficients.}\/
Indeed, this is an {\it extra}\/ enhancement which emerges {\it beyond}\/ the mere
effects of normalization, and which ultimately reflects the fact that the pseudoscalar
and axial-vector coefficients are locked together as a single degree of freedom
through a relation such as that in Eq.~(\ref{current}).
Or, phrased somewhat differently, 
the factor of $ 2 m_\psi$ which appears in Eq.~(\ref{current}) ---  and which ultimately leads directly to
our enhancement factor in Eq.~(\ref{eq:DeltaqPrimeEqs}), thereby driving the $\Delta \tilde q^{(N)}$ coefficients
above unity ---
follows not from a  normalization  but rather from an {\it equation of motion}\/.
Thus, our enhancement factor reflects far more than mere normalization conversion;
it is instead deeply rooted in the dynamics of the quark and nucleon fields and 
the fact that their corresponding pseudoscalar and axial-vector currents 
are tied together through Eq.~(\ref{current}).

\begin{table}
\begin{tabular}{||l | c  c||}
\hline
\hline
\multicolumn{2}{||c}{~~~~~~~~~~~~~~~~~$N=p$~} & ~~~~$N=n$~~~  \\ 
\hline
 ~~~$\Delta \tilde u^{(N)\phantom{^X}}$  &    $\phantom{-}110.55 \pm 21.87$  &        $-108.03 \pm 21.33$ \\
 ~~~$\Delta \tilde d^{(N)}$~     &         $-107.17 \pm 21.14$   &     $\phantom{-}108.60 \pm 21.29$    \\
 ~~~$\Delta \tilde s^{(N)}$~     &       ~$ -3.37 \pm 1.01$  &      ~$ -0.57 \pm 0.78$ \\         
 ~~~$\Delta  \tilde G^{(N)}$~           &      ~~~($395.2 \pm 124.4$)~MeV~~  &   ~~~($35.7 \pm 95.4$)~MeV~~ \\
\hline
\hline
\end{tabular}
\caption{ ~Numerical values for the pseudoscalar coefficients $\Delta \tilde q^{(N)}$, as
obtained from Eq.~(\ref{eq:DeltaqPrimeEqs}).  Details concerning the calculation of these
quantities and their associated uncertainties are discussed in the text.  It is readily observed
that these pseudoscalar coefficients $\Delta \tilde q^{(N)}$ are larger than the corresponding
axial-vector coefficients $\Delta q^{(N)}$ in Table~\protect\ref{tab:Delta} 
by a factor of ${\cal O}(10^2-10^3)$.  This can enhance the dark-matter/nucleon
scattering amplitudes associated with pseudoscalar interactions, and thereby potentially overcome the
velocity suppression that would otherwise render such cases unobservable in direct-detection experiments.}
\label{tab:deltaprime}
\end{table}

Using the algebraic results in Eq.~(\ref{eq:DeltaqPrimeEqs}) and the
numerical results in Table~\ref{tab:Delta}, 
we can evaluate the $\Delta \tilde q^{(N)}$ coefficients explicitly.
Our results, along with associated uncertainties, 
are shown in Table~\ref{tab:deltaprime}.
As we see, the pseudoscalar $\Delta \tilde q^{(N)}$ coefficients
are indeed larger than the corresponding axial-vector $\Delta q^{(N)}$ coefficients
in Table~\ref{tab:Delta} by a factor of ${\cal O}(10^2-10^3)$ in each case, as promised.
Indeed, as we shall demonstrate below, it is precisely the relatively large size of the 
pseudoscalar coefficients $\Delta \tilde q^{(N)}$ which
compensates for the velocity suppression.
For these numerical calculations, 
we have taken $\eta = 0.0 \pm 0.33$, as discussed above,
and we have taken the masses of the light quarks 
(and their associated uncertainties)
from Ref.~\cite{PDG}.
In particular, we have taken
$m_u=2.3 \pm 0.7$~MeV,
$m_d= 4.8 \pm 0.5$~MeV,
and
$m_s= 95 \pm 5$~MeV,
corresponding to the quark masses at the renormalization scale  $\mu = 2$~GeV
in the $\overline{\rm MS}$ renormalization scheme,
and then rescaled each mass and uncertainty by a factor of $1.35$ in order 
to account for the effect of renormalization-group running down to the scale 
$\mu \approx 1$~GeV appropriate for dark-matter/nucleon scattering~\cite{PDG}.
All uncertainties were then added together in quadrature in order to
produce the final uncertainties quoted in Table~\ref{tab:deltaprime}.

As evident from Table~\ref{tab:deltaprime},
the results for the pseudoscalar $\Delta \tilde q^{(N)}$ coefficients no longer respect
quark-level isospin invariance, as defined in Eq.~(\ref{quarklevelisospin}).
[In this connection we observe that quark-level isospin invariance would also require
$\Delta \tilde G^{(p)} = \Delta \tilde G^{(n)}$.]
This is a clear distinction relative to the axial-vector case in Table~\ref{tab:Delta},
but there are several ways in which to understand this result.
At an algebraic level, the breaking of quark-level isospin invariance
arises because the transition from the axial-vector coefficients
to the pseudoscalar coefficients
explicitly involves the quark masses;  by contrast, the axial-vector
coefficients were derived under approximations in which the light quarks
are effectively treated as massless.
Or, phrased somewhat differently, the leading terms in the axial-vector matrix
elements are independent of the quark masses;  it is only the subleading terms which
depend on these masses explicitly.
This is different from the situation one faces in dealing with the pseudoscalar
matrix elements, for which the leading terms are already mass dependent.
On a more physical level,
this difference can alternatively be understood
as arising from the fact that
the axial-vector current
is somewhat special in that its matrix element 
essentially counts the number of fermions minus anti-fermions,
weighted by chirality and normalized to the mass of the nucleon bound state.
[This is analogous to the vector-current matrix element, which
also counts the normalized number of fermions minus anti-fermions but
without a chirality weighting.]
As a result, the leading-order results in the axial-vector case depend on the
number and charges of the parton constituents, but not their masses.
This is to be contrasted with the pseudoscalar matrix elements,
for which an additional quark mass dependence can arise.
 
It should also be noted that while the uncertainties quoted in Table~\ref{tab:deltaprime}
are reliable in terms of their approximate overall magnitudes, there are certain
effects which we have not taken into account which might alter these results
slightly.   
Such effects will be discussed more fully
as part of an exhaustive uncertainty analysis in Ref.~\cite{student}.
For example, we have treated the uncertainties in Table~\ref{tab:Delta} for 
the axial-vector $\Delta q^{(N)}$ coefficients as independent of each other (\ie, uncorrelated), but 
in truth (see, \eg, Ref.~\cite{EllisOliveSavage:2008})
the $\Delta u^{(N)}$ and $\Delta d^{(N)}$ 
coefficients are actually extracted as linear combinations of
two more fundamental variables $a_3^{(N)}$ and $a_8^{(N)}$.  It is actually the uncertainties on these
latter variables which are independent, not those on the $\Delta q^{(N)}$ coefficients.
Likewise, the uncertainties on the quark masses 
are also not independent, as these masses are typically 
extracted in terms of a single reference quark mass (typically that of the down quark) and the ratios 
of the other quark masses relative to this reference mass.
The truly independent uncertainties are therefore those for the down-quark mass and the corresponding ratios.
Moreover, the uncertainties on the quark masses are not necessarily Gaussian, since they
typically have both systematic and random contributions.
Combining these into a single uncertainty, as we have done here, and then treating this single uncertainty
as Gaussian when performing a quadrature-based analysis represents yet another approximation.
Indeed, $\eta$ is an example of a variable whose uncertainty is completely systematic rather than
experimental, yet its uncertainty is being treated as if were Gaussian as well.
Finally, there is even some leeway concerning how one treats isospin symmetry 
in a rigorous uncertainty analysis.  Isospin symmetry, as mentioned above, is usually invoked in 
order to relate quantities such as $\Delta u^{(p)}$ and $\Delta d^{(n)}$ --- indeed, it is typically
the case that these quantities
are not measured independently.  As a result of this presumed isospin symmetry, 
these quantities are necessarily
quoted as having the same central values and same
quoted uncertainties, as indicated in Table~\ref{tab:Delta}.  
However, it is not clear whether these uncertainties should be treated as
independent or correlated when performing a quadrature-based uncertainty analysis of the sort we
are performing here.  While isospin symmetry would dictate that these uncertainties be treated as
completely correlated, we know that isospin symmetry is only approximate in nature.  Indeed,
as mentioned above,
the results in Table~\ref{tab:deltaprime}
for the central values of our pseudoscalar $\Delta \tilde q^{(N)}$ 
coefficients already fail to respect isospin symmetry because of
their explicit dependence on the light-quark masses.
We have therefore
opted to treat the uncertainties in Table~\ref{tab:Delta} as completely independent and uncorrelated.

Despite these observations, the uncertainties quoted in Table~\ref{tab:deltaprime} are
correct in terms of their overall magnitudes.
It is also evident that the pseudoscalar uncertainties quoted in Table~\ref{tab:deltaprime}
are somewhat larger, in relative terms, than 
the corresponding axial-vector uncertainties quoted in Table~\ref{tab:Delta}.
This is partially due to the dependence of the pseudoscalar results on a constraint which stems from a large-$N_c$
approximation.  As a result of these larger uncertainties, 
we see that certain quantities in Table~\ref{tab:deltaprime}, such as 
$\Delta \tilde s^{(n)}$
and $\Delta \tilde G^{(n)}$,
are actually consistent with zero.
As we shall see, these results will
lead to considerably larger uncertainties
for our eventual pseudoscalar dark-matter/nucleon
couplings.

Finally, we now turn to the pseudoscalar $\Delta \tilde q^{(N)}$ 
coefficients for the heavy quarks $q=Q\equiv c,b,t$.
As we shall see, these quantities will be 
relevant if our dark matter couples to such quarks.
In the axial-vector case, the analogous coefficients 
were taken to be zero, reflecting the fact that such quarks are heavy
and make only negligible contributions to 
axial-vector couplings.
For pseudoscalar couplings, by contrast, the situation is different.
Because of the current-algebra relation in Eq.~(\ref{current}), we
see that the pseudoscalar current is related to the {\it derivative}\/  
of a different current involving the same heavy fermions.
However, if the fermions in question are sufficiently heavy, they
will have no dynamics and this derivative must vanish.
We thus obtain the relation
\beq
  2 m_Q \langle N_f|  \overline{Q} i\gamma^5 Q |N_i\rangle ~=~
         - {\alpha_s\over 4\pi} \langle N_f| G\tilde G| N_i\rangle~,
\eeq
from which we see that
\beqn
   \langle N_f|  \overline{Q} i\gamma^5 Q |N_i\rangle &=&
         - {1\over m_Q} \, {\alpha_s\over 8\pi} \langle N_f| G\tilde G| N_i\rangle\nonumber\\
   &=&  -{1\over m_Q} \Delta \tilde G^{(N)} 
   \langle N_f|  \overline{N} i\gamma^5 N |N_i\rangle ~,~~~~~~~~~
\eeqn
where the values of $\Delta \tilde G^{(N)}$ are given algebraically 
in Eq.~(\ref{eq:DeltaqPrimeEqs}).
We thus find that
\beq
    \Delta \tilde Q^{(N)} ~=~  - {1\over m_Q} \Delta \tilde G^{(N)}~.
\eeq


\subsection{Pseudoscalar dark-matter/nucleon couplings and the effects of isospin violation
\label{CouplingScenarios}}

We now turn to the actual quantities $g_{\chi N}$
which parametrize how the dark-matter fermion $\chi$ couples to nucleons $N$ in the case of 
pseudoscalar interactions.
As evident in Eq.~(\ref{gchincouplings}),
these effective couplings $g_{\chi N}$ are directly determined in terms of the 
$\Delta \tilde q^{(N)}$ coefficients for both light and heavy quarks: 
\begin{equation}
  g_{\chi N} ~=~ \sum_{q=u,d,s}c_{q}\Delta \tilde q^{(N)}-
    \sum_{Q=c,b,t}\frac{c_{Q}}{m_Q}\Delta \tilde G^{(N)}~,
  \label{eq:gchiNgeneral}
\end{equation}
where the numerical values of the $\Delta \tilde q^{(N)}$ 
and $\Delta \tilde G^{(N)}$ coefficients 
are listed in Table~\ref{tab:deltaprime}.

The only task remaining, then, is to determine the values for the quark couplings $c_q$ 
(henceforth taken to collectively denote the couplings for both light and heavy quarks).
Of course, the expression in Eq.~(\ref{eq:gchiNgeneral}) for the $g_{\chi N}$ is completely
general and applicable for any choice of operator coefficients $c_q$ in the fundamental theory.
In principle, any assignment of the $c_q$ consistent with phenomenological
constraints is therefore permitted.  
However, for concreteness, in this paper we shall focus primarily on
three particular benchmark scenarios:

\begin{itemize}

\item {\it Scenario~I}\/:~~  The case in which the $c_q$ for all up-type quarks take a
  common value $c_u = c_c = c_t$ and the $c_q$ for all down-type quarks likewise 
  take a (potentially different) common value $c_d = c_s = c_b$.  For this scenario,
  we parametrize these two independent operator coefficients in terms of a mass scale $M_{\rm I}$ 
  and an angle $\theta$ such that $c_u / \Lambda^2 = \cos\theta / M_{\rm I}^2$ 
  and $c_d / \Lambda^2 = \sin\theta / M_{\rm I}^2$.    
  It then follows that $\tan \theta = c_d/c_u$ and $M_{\rm I}^2 = \Lambda^2/\sqrt{c_u^2 + c_d^2} $.
  Note that for $\theta=\pi/4$,  this coupling structure respects quark-level isospin invariance.
  Varying $\theta$ will thus allow us to study the effects of isospin violation in a continuous fashion.

\item {\it Scenario~II}\/:~~ A generalization of the oft-studied case in which the $c_q$
  are proportional to the Yukawa couplings $y_q$ between the quarks and the SM
  Higgs boson, and thus to $m_q$.  This scenario is motivated by the
  minimal-flavor-violation (MFV) assumption that the quark Yukawa couplings are
   wholly responsible for flavor violations.
  The generalization we consider here is one in which the $c_q$ for the up-type quarks may
  also be scaled by an overall multiplicative factor relative to the $c_q$ for the 
  down-type quarks.  
  Specifically, for this scenario, we define a mass scale $M_{\rm II}$ and an angle 
  $\theta$ such that $c_q/\Lambda^2 = m_q \cos\theta / M_{\rm II}^3$
  for up-type quarks and $c_q/\Lambda^2 = m_q \sin\theta / M_{\rm II}^3$ for 
  down-type quarks.  
  It then follows that $\tan \theta = (c_d m_u) / (c_u m_d)$ and 
   $M_{\rm II}^3 = \Lambda^2/\sqrt{(c_u/m_u)^2 + (c_d/m_d)^2}$, where
    $c_u/m_u = c_c/m_c = c_t/m_t$ and $c_d/m_d = c_s/m_s = c_b/m_b$. 

\item {\it Scenario~III}\/:~~ The related case in which the $c_q$  
  are non-vanishing only for the first-generation quarks --- \ie, in which
  $c_u$ and $c_d$ are arbitrary, but in which $c_s = c_c = c_b = c_t = 0$.
  For this scenario, we likewise define $M_{\rm III}$ and $\theta$ such that
  $c_u /\Lambda^2 = m_u \cos\theta / M_{\rm III}^3$ and 
  $c_d /\Lambda^2 = m_d \sin\theta / M_{\rm III}^3$.
  This coupling structure is of particular interest from a direct-detection perspective,
   implying that $c_u$ and $c_d$ uniquely determine the 
  effective dark-matter/nucleon couplings $g_{\chi p}$ and $g_{\chi n}$ 
  in Eq.~(\ref{eq:gchiN}), and {\it vice versa}\/.  
  Moreover, with the couplings for the second- and third-generation quarks 
   set to zero, this scenario is the only one which does not involve couplings 
   which are essentially irrelevant for direct detection.
   Since non-zero couplings for second- and third-generation quarks 
   could potentially 
    have a significant effect on the rates for dark-matter production at colliders~\cite{monob},
    this scenario is therefore in some sense the most ``conservative'' in that it 
    does not assume any channels which might enhance collider signatures without affecting
    direct-detection signals.
    Study of this scenario will therefore  
    lead to the most conservative set of limits consistent with collider data.

\end{itemize}

We emphasize that these three scenarios represent physically distinct coupling
structures between $\chi$ and the SM quarks.
It is for this reason that each scenario has been associated with its own
independent mass scale above.

Given these three scenarios, we can now proceed to examine the behavior of
our pseudoscalar dark-matter/nucleon couplings as functions of $\theta$ in each scenario.
For Scenario~I, the results in Table~\ref{tab:deltaprime} 
yield the effective pseudoscalar couplings  
\begin{align}
  g_{\chi p}& ~=~ ~~~110.2\cos \theta -110.6 \sin{\theta} \nonumber \\
  g_{\chi n}& ~=~ -108.1\cos \theta + 108.0\sin{\theta}~.
\label{eq:gchiN}
\end{align}
Likewise, given the uncertainties in Table~\ref{tab:deltaprime}, we find that
the associated {\it uncertainties}\/ in these couplings
are given by rather complicated expressions which can be extremely well approximated
as
\beqn
 \delta g_{\chi p}  &\approx & \big |21.79 \cos\theta - 21.88\sin\theta\big |\nonumber\\
 \delta g_{\chi n}  &\approx & \big |21.32 \cos\theta - 21.33\sin\theta\big |~.
\label{uncerts}
\eeqn
We immediately note that both the couplings {\it and}\/ their associated uncertainties
are nearly vanishing at the quark-level isospin-preserving point $\theta=\pi/4$.
Alternatively, given the couplings in Eq.~(\ref{eq:gchiN}), we can solve 
for the value $\theta^\ast$ at which {\it nucleon-level}\/ isospin preservation takes
place --- \ie, the value $\theta^\ast$ at which $g_{\chi p} = g_{\chi n}$.
We find that in this scenario, the nucleon-level isospin-preserving point is extremely 
close to the quark-level isospin-preserving point, with only a very small net displacement
$\theta^\ast - \pi/4 \approx  -8.45 \times 10^{-4}$~radians.

At the nucleon-level isospin-preserving point, we find that $g_{\chi p}=g_{\chi n} \approx -0.155
\pm 0.25$ --- a value consistent with zero.  
This is remarkable, representing a situation in which dark matter couples to quarks,
but not to nucleons!
Moreover, this is to be compared with the couplings that emerge
for other, isospin-violating values of $\theta$.  For example, we find that the proton coupling
takes the value $|g_{\chi p}|\approx 110.6 \pm 21.9$ 
at $\theta=\pi/2$,
and reaches a maximum value $|g_{\chi p}|\approx 156.2 \pm 30.9$ at $\theta\approx 3\pi/4$.
The behavior of the neutron coupling $|g_{\chi n}|$ is similar.
Thus, relative to the central values of these couplings at the isospin-preserving point $\theta=\theta^\ast$,
we see that these couplings experience a huge enhancement which can grow as large as a factor
of $10^3$!

It is important to understand the physical origin of the cancellation of
this coupling in the isospin-preserving case.
Ultimately, this cancellation is the direct result of
the fact that we (like most researchers in this field) 
are working in the large-$N_c$ chiral limit
in which relations such as that in Eq.~(\ref{largeNc}) apply.
Indeed, since the strange-quark contribution in Eq.~(\ref{largeNc})  
is small, this relation immediately implies that 
the isospin-conserving case will experience a cancellation.
At a physical level, this can equivalently be understood as follows.
In general, one can consider an approximation in which dark-matter/nucleon scattering is
considered to be mediated by neutral-meson exchange.  
In this approximation, the dominant contribution is
from pion exchange, and the pion couples to the first-generation quarks in a way which is
maximally isospin-violating.  (By contrast, the isospin-{\it conserving}\/ 
case would involve the $\eta$ and $\eta'$ states as mediators, but these states are
much heavier than the pion.)
Indeed, Eq.~(\ref{largeNc}) emerges
in the large-$N_c$ chiral limit 
precisely because this is the limit in which the $\eta'$ decouples.
Nevertheless, we have also explicitly taken into account possible
small departures from the large-$N_c$ limit when we introduce our $\eta$-dependent
``error'' term in Eq.~(\ref{largeNc2}).
The fact that the proton- and neutron-couplings continue to vanish ---
even within the resulting uncertainties --- demonstrates that this cancellation
is robust against small departures from the large-$N_c$ limit.

We therefore conclude that isospin violation in Scenario~I produces
a huge enhancement in the corresponding
pseudoscalar proton and neutron couplings.  This is the direct result of 
the relatively large pseudoscalar coefficients
$\Delta \tilde q^{(N)}$ in Table~\ref{tab:deltaprime}, operating within the 
framework of the particular quark coupling structure associated with Scenario~I.~
However, it is important to stress that there is nothing intrinsic to the coupling 
structure of Scenario~I by itself which  causes such large proton and nucleon couplings to emerge.
For example, as an algebraic exercise, 
we can calculate the proton and neutron couplings
that would emerge under Scenario~I  
in the {\it axial-vector}\/ case --- \ie, using the axial-vector coefficients $\Delta q^{(N)}$
in Table~\ref{tab:Delta} rather than the pseudoscalar coefficients $\Delta \tilde q^{(N)}$
in Table~\ref{tab:deltaprime}. 
In this case, because of the fact that isospin symmetry is exactly preserved for the $\Delta q^{(N)}$ coefficients,
both quark-level and nucleon-level isospin preservation coincide exactly at $\theta =\pi/4$.
Indeed, at this point we find
$g_{\chi p} = g_{\chi n} \approx 0.303 \pm 0.12$,
while the maximum value taken by these couplings for any isospin-violating value of 
$\theta$ is $|g_{\chi p}|\approx 0.865\pm 0.15$ at $\theta\approx 2.714$ and $g_{\chi n}\approx 0.812 \pm 0.15$ 
at $\theta\approx 1.974$.
Thus, for the axial-vector case,
we see that isospin violation is capable of increasing the proton and neutron couplings only by
mere factors of $2.85$ and $2.68$ respectively.

We also note, of course, that the overall {\it scale}\/ of the axial-vector couplings 
is significantly smaller than that for the pseudoscalar couplings.
While it is perhaps inappropriate to compare the magnitudes of these different couplings against each
other (because they correspond to different operators with gamma-matrix bilinears  
exhibiting entirely different tensorial properties), at a purely algebraic level this difference can once again
be attributed to the larger values of the $\Delta \tilde q^{(N)}$ coefficients that enter  
the calculation of the pseudoscalar proton and neutron couplings as compared with the
values of the 
$\Delta  q^{(N)}$ coefficients that enter  
the calculation of their axial-vector counterparts.

\begin{figure*}[t]
\begin{center}
  \epsfxsize 2.15 truein \epsfbox {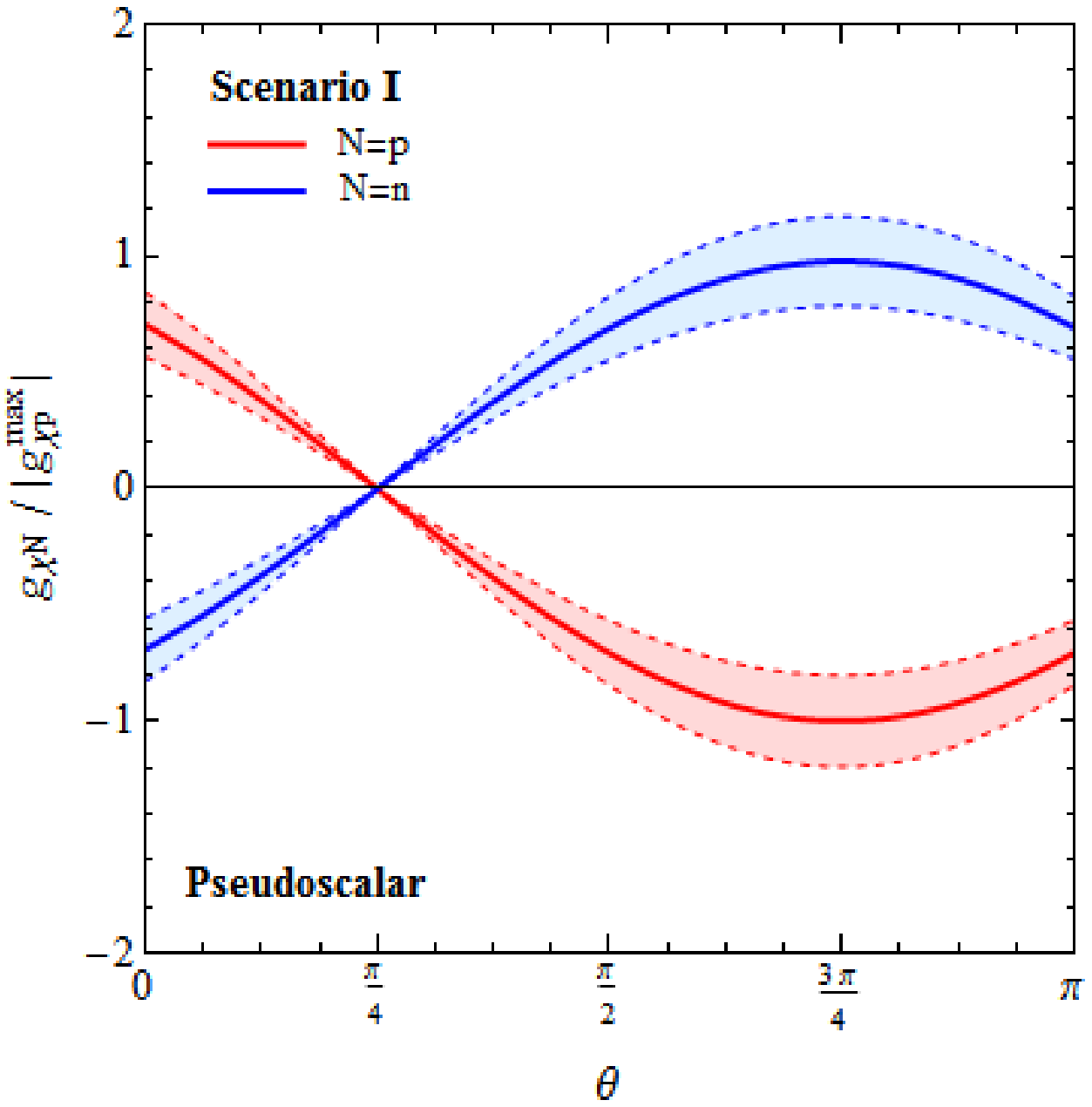}~~
  \epsfxsize 2.15 truein \epsfbox {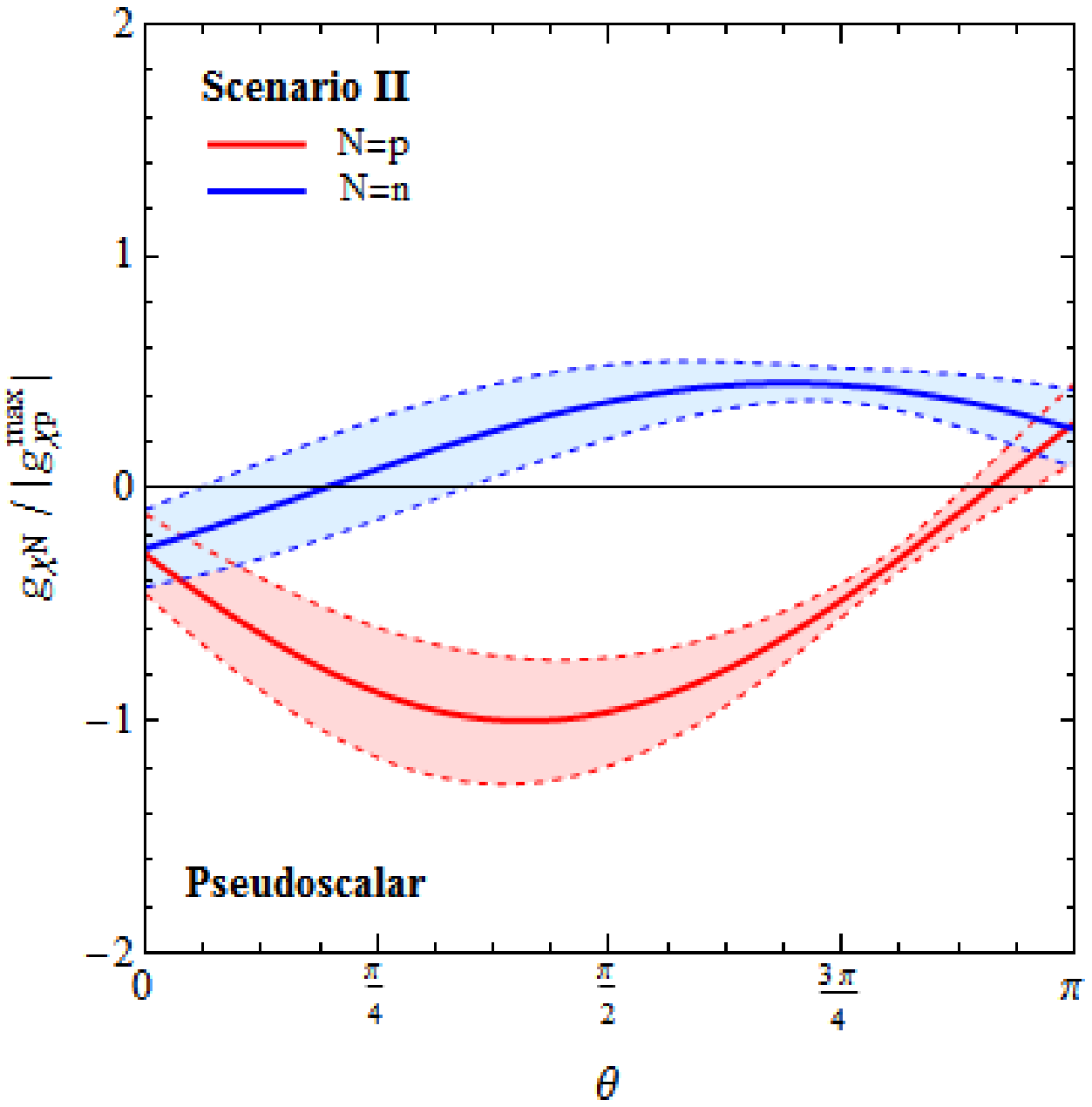}~~
  \epsfxsize 2.15 truein \epsfbox {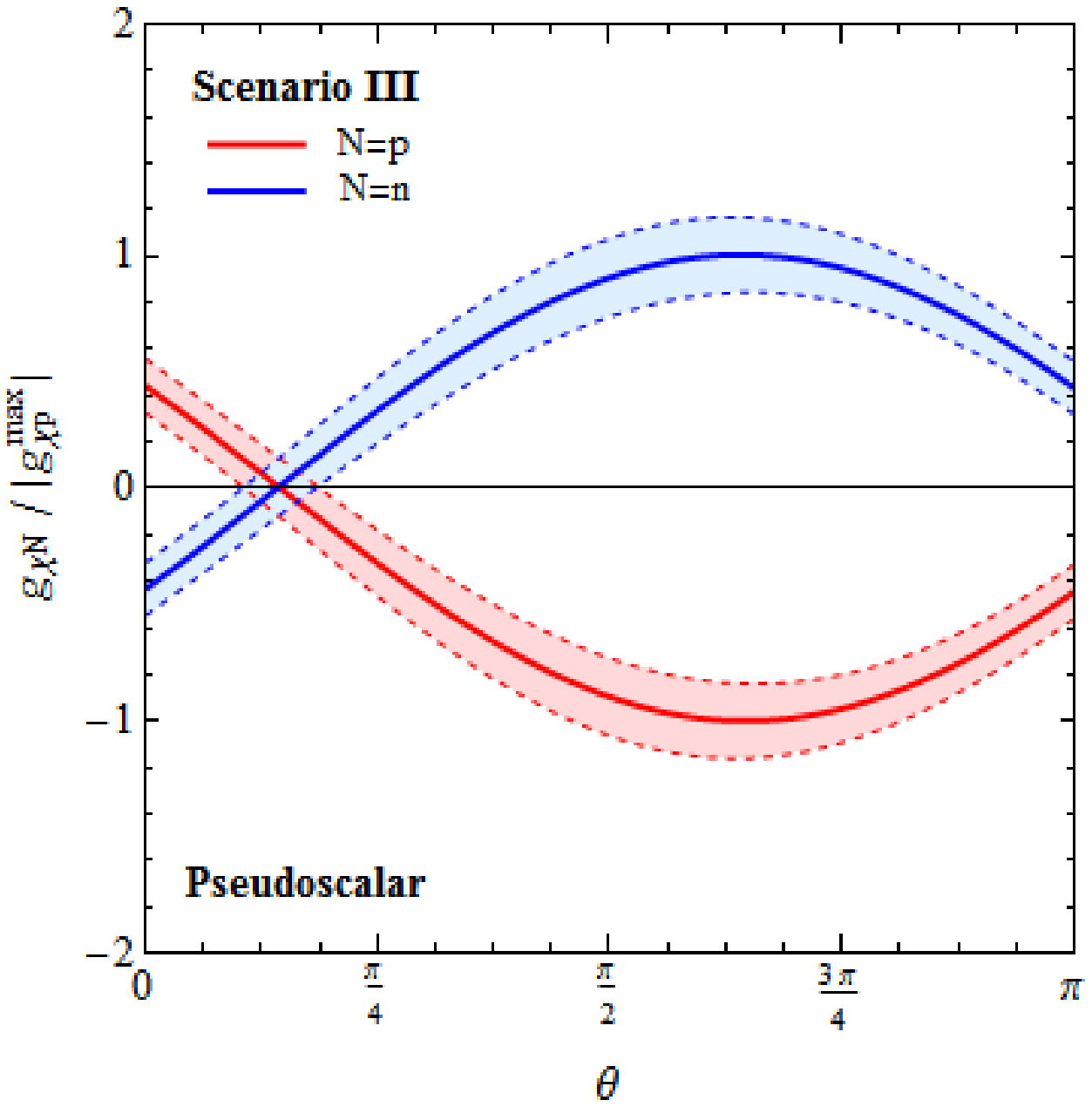}
\end{center}
  \begin{center}
  \raisebox{1.0 truein}{ \parbox[r]{1.0 truein}{\sl for comparison\\ \hfil ~~~~purposes:}}~~~~
    \epsfxsize 1.6 truein \epsfbox {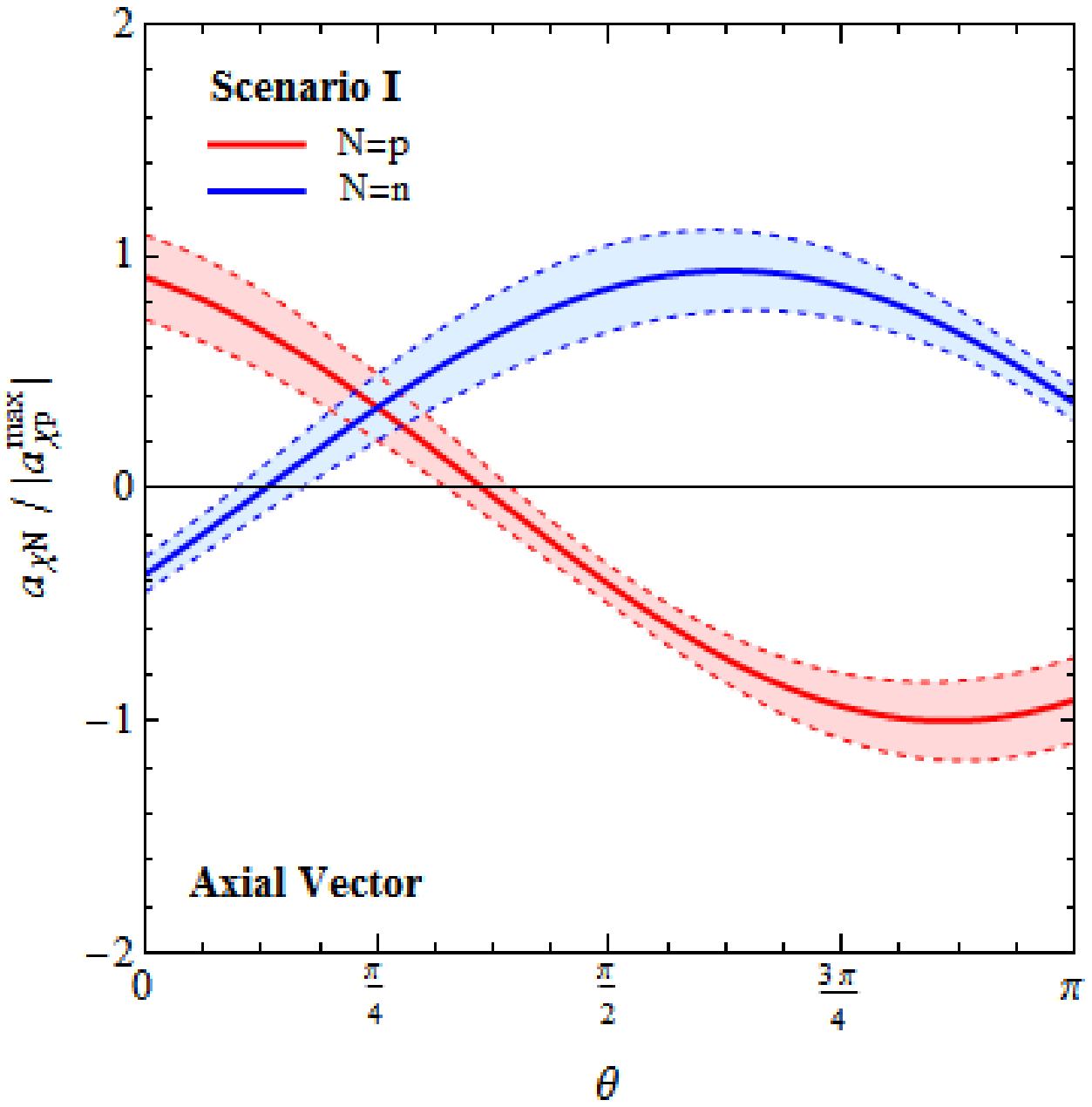}~
    \epsfxsize 1.6 truein \epsfbox {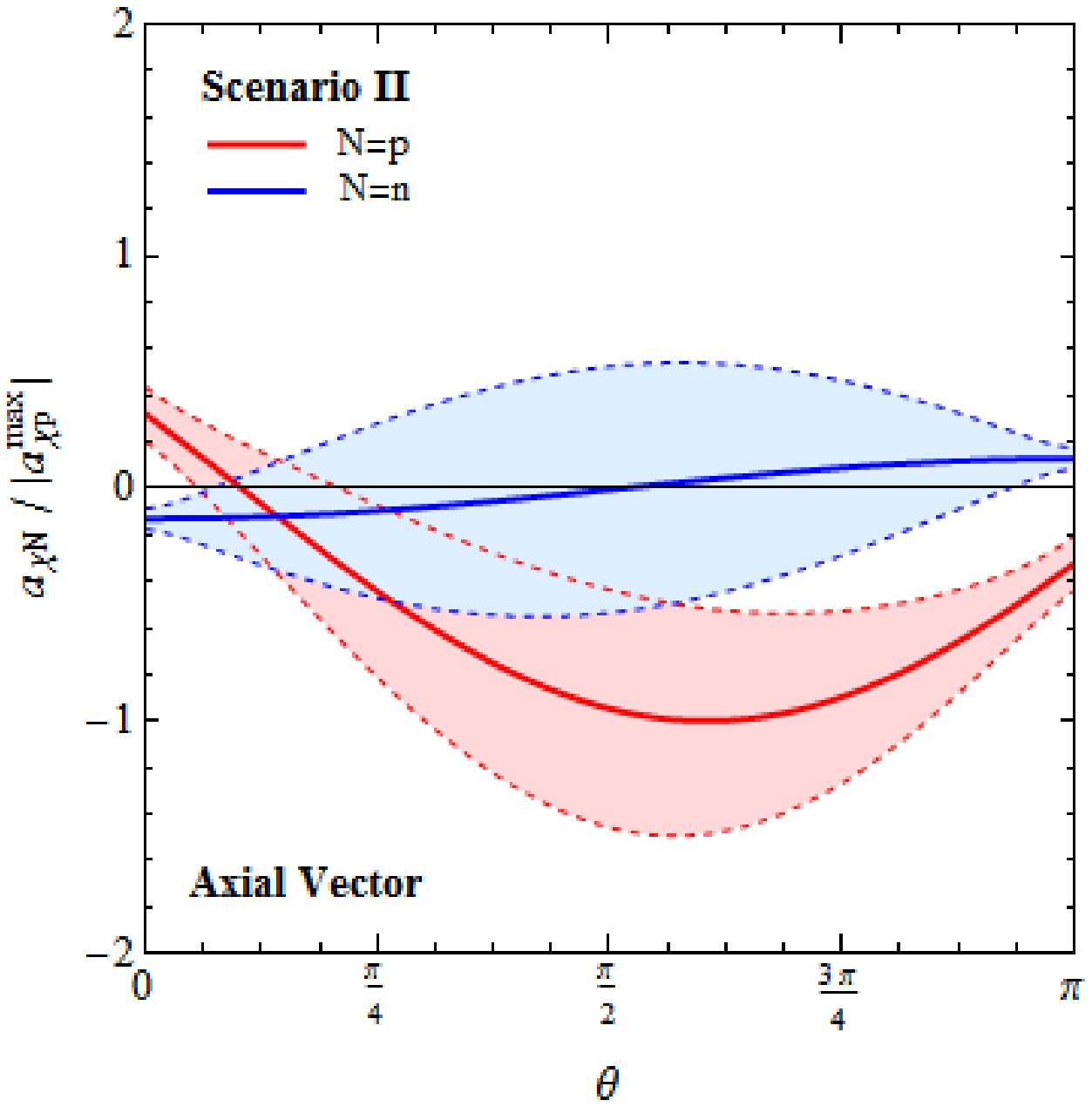}~
    \epsfxsize 1.6 truein \epsfbox {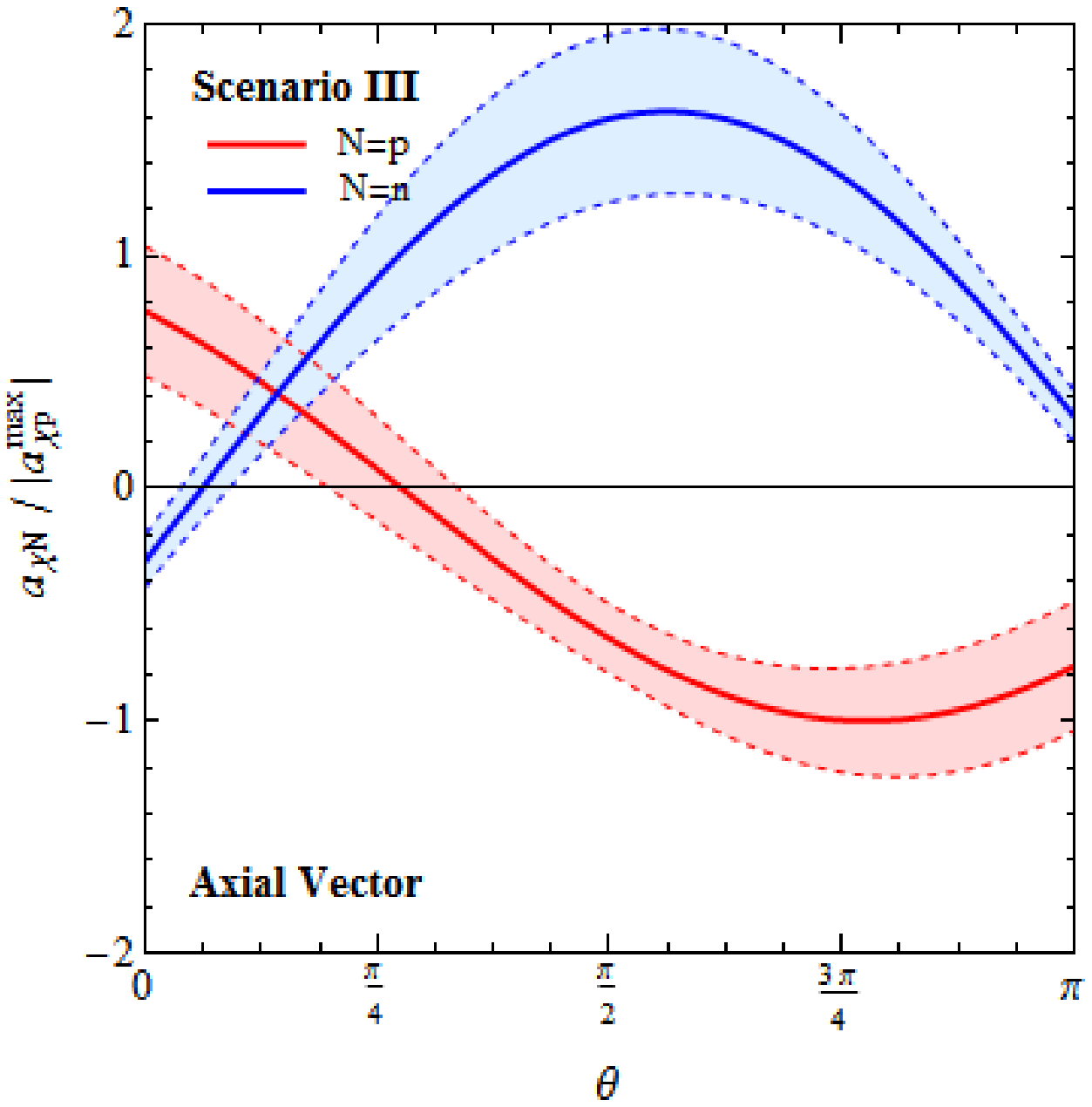}
  \end{center}
\caption{The effective proton and neutron dark-matter couplings $g_{\chi p}$ (red) and $g_{\chi n}$ (blue),
plotted as functions of $\theta$ for each of the three coupling scenarios discussed in the text.  
Panels in the upper row show the behavior of the pseudoscalar couplings in each scenario,
while the panels in the lower row show the behavior of the corresponding axial-vector couplings.
The dashed lines in each panel correspond to the central values for these couplings, while the shaded 
regions indicate the $1\sigma$ uncertainty bands around these central values.  Note that 
in each panel, both $g_{\chi p}$ and $g_{\chi n}$ have been normalized to the maximum 
possible central value of $|g_{\chi p}|$ attainable in each scenario.}
\label{fig:gchiNPlot}
\end{figure*}

In Fig.~\ref{fig:gchiNPlot}, 
we have plotted the pseudoscalar proton and neutron couplings $g_{\chi p}$ and
$g_{\chi n}$,
along with their corresponding uncertainties,
as functions of $\theta$ for all three of our coupling scenarios. 
For comparison purposes, we have also plotted the corresponding axial-vector couplings
as functions of the same variable $\theta$. 
Moreover, in each case we have normalized the proton and neutron couplings to the maximum value 
that the proton coupling ever attains as a function of $\theta$.

Many features of these plots are worthy of note.
Focusing first on the pseudoscalar couplings,
we have already remarked
that a significant degree of cancellation occurs within Scenario~I
when isospin is conserved at the nucleon level:  both the proton and neutron pseudoscalar
couplings, along with their associated uncertainties, become extremely small
as a result of a near-perfect cancellation between their individual up-quark and down-quark contributions.
As remarked earlier, this is then a situation in which our dark matter couples to quarks, but not to
nucleons! 
What is now apparent from Fig.~\ref{fig:gchiNPlot}, however, is that this cancellation is a relatively
sharp one, and that any movement away from this isospin-conserving value of $\theta$ in either direction
results in a significant enhancement of these pseudoscalar nucleon couplings.
As indicated above, 
this results in an $\mathcal{O}(10^3)$ enhancement in the pseudoscalar couplings
for isospin-violating scenarios relative to the na\"{i}ve isospin-conserving case, and 
thus to an $\mathcal{O}(10^6)$ enhancement in the cross-section for the scattering
of $\chi$ off atomic nuclei.  Thus, we see that even a relatively small amount of isospin 
violation can have a dramatic effect on direct-detection rates!  
 
The above behavior occurs for Scenario~I.~
However, we now see from Fig.~\ref{fig:gchiNPlot} that similar behavior also occurs for Scenario~III,
albeit at a somewhat shifted value of $\theta$.
This feature is also easy to understand.
In Scenario~I, the cancellation that occurs at $\theta^\ast$ 
truly reflects an approximate isospin symmetry.  Indeed, while
the term in Eq.~(\ref{eq:gchiNgeneral}) proportional to $\Delta \tilde G^{(N)}$ is manifestly
isospin-violating, this contribution is suppressed by several orders of magnitude compared to the 
contributions from the light quarks in this scenario.  Moreover, since the $c_q$ are independent of the 
quark masses in Scenario~I and since the $\Delta \tilde q^{(N)}$ are {\it approximately}\/ isospin-conserving
(particularly for the two lightest quarks),
this cancellation occurs for a value of $\theta^\ast$ very close to $\theta \approx \pi/4$.  
Of course, in Scenario~III, the $\Delta \tilde G^{(N)}$ contribution to the couplings vanishes outright because the 
dark-matter particle does not couple to the heavy quarks.  However, the above cancellation now occurs 
at the value $\theta^\ast = \tan^{-1}(m_u/m_d)\approx 0.45$ rather than at $\theta^\ast \approx \pi/4$, 
for within Scenario~III it is only at this shifted angle that $c_u = c_d$.
Furthermore, within Scenario~III, we see that the uncertainties are no smaller at $\theta^\ast$ than they are 
at any other angle --- another distinction relative to Scenario~I.~

Finally, we observe that the pseudoscalar couplings shown for Scenario~II 
differ quite significantly from those shown for Scenario~III, both in terms
of the locations of the nucleon-level isospin-preserving points as well as the 
overall magnitudes of the associated uncertainties.  
These differences ultimately reflect 
the contributions from the second- and third-generation quarks.
One notable feature in Scenario~II, for example,
is the fact that the sort of cancellation which occurs for Scenarios~I and III
does not occur for Scenario~II.~
The reason for this is also easy to understand.
In Scenario~II, we have $c_q \propto m_q$ for all quark species.  For such a coupling structure, 
it turns out that the magnitudes of the two terms on the right side of Eq.~(\ref{eq:gchiNgeneral})
are roughly commensurate.  Thus, even if $\theta$ were set at a value for which
the light-quark contributions roughly cancelled, the heavy-quark contributions would still be
significant.  Indeed, for this scenario, we find that
nucleon-level isospin preservation 
arises at $\theta^\ast \approx 3.12$ --- a value much closer to $\pi$ than to
$\pi/4$ --- but the proton and neutron couplings at this point are clearly non-zero.

In Fig.~\ref{fig:gchiNPlot} we have also illustrated what occurs 
for the corresponding {\it axial-vector}\/ couplings in each scenario.
For example, as already discussed above, we see that the isospin-preserving
points no longer correspond to vanishing proton and neutron couplings --- even
for Scenarios~I and III.~  
Thus isospin violation
will no longer produce as dramatic an enhancement for the 
axial-vector proton and neutron couplings as it does for the 
corresponding pseudoscalar couplings, even in these scenarios.
Moreover, we observe that unlike the situation for the pseudoscalar couplings,
there are no values of $\theta$ in Scenarios~I or III for which both $g_{\chi p}$ and $g_{\chi n}$ 
vanish simultaneously.
Thus, for axial-vector couplings, 
dark-matter couplings to quarks always imply a dark-matter coupling to at least one nucleon.
Furthermore, 
we see that the uncertainties are so large for the axial-vector neutron coupling
in Scenario~II that the value of this coupling is consistent with zero for almost all values of $\theta$.
Finally, although it is not visible from the plots in Fig.~\ref{fig:gchiNPlot},
we again stress that the overall magnitude of the axial-vector couplings
is a factor of ${\cal O}(10^2 - 10^3)$ smaller than the magnitude of the
pseudoscalar couplings.
This is perhaps the most important difference of all.

Despite the rather compelling nature of these differences,
it is important to bear in mind that 
the pseudoscalar and axial-vector couplings correspond to entirely
different operators.  Thus, a direct comparison between these couplings
is fraught with a number of theoretical subtleties.
For example, Scenarios~II and III are rather unnatural within an axial-vector framework, and it is difficult
to imagine a high-scale model which might yield such an axial-vector effective operator
with the quark-level couplings of Scenarios~II or III at lower energies.
This is completely different from what happens within the pseudoscalar framework, where the
coupling structures of Scenarios~II and III  are particularly well motivated.
Nevertheless, we have undertaken such a direct coupling-to-coupling comparison 
in order to expose the primary numerical differences that emerge 
when the axial-vector $\Delta q^{(N)}$ coefficients of Table~\ref{tab:Delta}
are replaced with the 
pseudoscalar $\Delta \tilde q^{(N)}$ coefficients of Table~\ref{tab:deltaprime}.
Indeed, from a purely bottom-up perspective, the coupling structures of all three scenarios
can be taken to represent interesting benchmarks which are introduced purely for the purpose of 
studying varying resulting phenomenologies in a model-independent framework.
We have therefore chosen to study the resulting couplings 
free of any theoretical prejudice stemming from considerations of high-scale physics.

Of course, what ultimately matters in each case are not the couplings themselves,
but rather the implications of these couplings
for the reach of actual direct-detection experiments.
For example, we have seen that even a small amount of isospin 
violation can dramatically enhance our pseudoscalar couplings,
but it remains to be seen whether this effect
is large enough  to compensate for the 
velocity suppression which is also associated with pseudoscalar interactions,
and thereby render such interactions
potentially relevant for detection at the next generation 
of spin-dependent dark-matter direct-detection experiments. 
This is therefore the topic to which we now turn.


\section{CP or not CP, that is the question:  
An interlude on the choice of Lagrangian operators}


In this paper, our analysis has focused on those interactions between
dark matter and Standard-Model matter which  
take the form of effective four-fermi contact interactions
whose operators 
exhibit the double-bilinear form in Eq.~(\ref{operator}).
Thus far, our interest has focused on the unique
physics that emerges from assuming a pseudoscalar structure for the quark bilinear
in Eq.~(\ref{operator}), and indeed all of our results thus far have relied
on this choice.
However, we have yet to select a tensor structure for the corresponding
dark-matter bilinear, and 
Lorentz invariance dictates that there are only two possible choices open
to us:
\beqn
      {\cal O}_{\chi q}^{\rm (SP)} &\equiv & {c_q\over \Lambda^2} \, (\chibar \chi)\, (\qbar i\gamma^5 q)\nonumber\\
      {\cal O}_{\chi q}^{\rm (PP)} &\equiv & {c_q\over \Lambda^2} \, (\chibar i\gamma^5 \chi)\, (\qbar i\gamma^5 q)~.
\label{Lagchoices}
\eeqn
The first of these operators breaks CP symmetry,
while the second preserves it.
Unfortunately, we can proceed no further in our discussion of actual
direct-detection experimental prospects without making a specific choice 
between these two operators.
 
The CP-violating operator ${\cal O}_{\chi q}^{\rm (SP)}$ is often neglected in 
direct-detection studies, even in comparison with ${\cal O}_{\chi q}^{\rm (PP)}$.
One reason for this is that ${\cal O}_{\chi q}^{\rm (PP)}$ is CP-invariant 
and can therefore be generated at 
a non-trivial level in many top-down theoretical constructions which yield a stable 
dark-matter candidate, such as the constrained minimal supersymmetric model 
(CMSSM) in which there is no additional source of CP violation.
However, in a 
bottom-up effective-theory approach such as the one we adopt here, the aim is
to examine and constrain the properties of all possible interactions which 
could arise between the dark-matter candidate and the particles of the SM in as 
model-independent a framework as possible, without theoretical prejudice.  Indeed,
while the operator ${\cal O}_{\chi q}^{\rm (PP)}$ is typically assumed to be irrelevant for 
direct detection, it is instructive to revisit why this is the case --- and also why 
this is {\it not}\/ the case for ${\cal O}_{\chi q}^{\rm (SP)}$, despite the fact that the 
structure of the quark bilinear is the same in both cases.    

Let us first consider the situation in which $\chi$ couples to SM particles 
primarily via ${\cal O}_{\chi q}^{\rm (PP)}$.  We assume for the purposes of this discussion
that this operator provides the dominant contribution both to the cross-section
for nuclear scattering events at direct-detection experiments and to the annihilation 
rate of $\chi$ and $\overline{\chi}$ in the early universe. 
For purposes of illustration, we also restrict our attention
to the case in which $\chi$ couples to only one quark flavor;  thus only 
one of the $c_q$ is non-vanishing.  
We have already seen for ${\cal O}_{\chi q}^{\rm (PP)}$ that
both the dark-matter bilinear and the quark-bilinear give rise to a velocity
suppression in the dark-matter/nucleon cross-section for 
direct detection.  Thus, for ${\cal O}_{\chi q}^{\rm (PP)}$, the resulting
(spin-dependent) cross-section can be expected to scale like   
\begin{equation}
 {\rm PP:}~~~~~~  \sigma_{\mathrm{SD}}^{(\chi N)} ~\sim~ 
    \frac{c_q^2 \left[\Delta \tilde q^{(N)}\right]^2\mu_{\chi N}^6 v^4}{m_\chi^2 m_N^2 \Lambda^4}~,
\label{PP1}
\end{equation}
where $\mu_{\chi N} \equiv m_\chi m_N / (m_\chi + m_N)$ denotes the 
reduced mass of the $\chi$/nucleon system.  

There are clearly many unknown parameters in Eq.~(\ref{PP1}), making
it difficult to provide an actual numerical estimate of this cross-section.
However, we may appeal to a somewhat orthogonal constraint which
applies to any thermal dark-matter candidate:
that through which the annihilation rate of $\chi$ and $\chibar$ sets an overall
dark-matter abundance in the early universe. 
For ${\cal O}_{\chi q}^{\rm (PP)}$, the annihilation of
$\chi$ and $\overline{\chi}$ in the early universe has no chirality suppression  
since the initial state is CP-odd, with quantum 
numbers $S=0$, $L=0$, and $J=0$~\cite{Kumar:2013iva}.  In an $s$-wave annihilation
scenario of this sort, the thermal annihilation cross-section 
$\langle\sigma |v|\rangle$ scales like 
\begin{equation}
{\rm PP:}~~~~~~ \langle\sigma |v|\rangle ~\sim~ \frac{c_q^2 m_\chi^2}{\Lambda^4}
  \label{eq:ScalingOfAnnXSecLPP}
\end{equation}
at around the time of freeze-out.  Moreover, in order for the relic-abundance 
contribution from freeze-out to agree with observation (\ie, $\Omega_\chi \approx \OmegaDM$), 
this cross-section must be roughly $\langle\sigma |v|\rangle \sim 1$~pb at such
times.  

Given this constraint, we can substitute back into Eq.~(\ref{PP1})
in order to find that
\begin{equation}
    {\rm PP:}~~~~~~
  \sigma_{\mathrm{SD}}^{(\chi N)} ~\sim~ \left(1\mathrm{~pb}\right) \times
  \frac{\left[\Delta \tilde q^{(N)}\right]^2 \mu_{\chi N}^6 v^4}{m_\chi^4 m_N^2}~. 
  \label{eq:LPPv4Suppression}
\end{equation}
Since $v^4 \sim \mathcal{O}(10^{-12})$, we see that extremely 
large values of $\Delta \tilde q^{(N)}$ would 
be required to overcome this velocity suppression and yield a $\chi$/nucleon 
cross-section of sufficient magnitude to be probed at any foreseeable direct-detection 
experiment, even for low-mass dark matter.  Indeed, since both $\langle\sigma |v|\rangle$ 
and $\sigma_{\mathrm{SD}}^{(\chi N)}$ depend on $\Lambda$ in the same manner for
a thermal relic, this unhappy consequence exists regardless of the scale $\Lambda$ 
at which $\Omega_\chi$ is generated via thermal freeze-out for a dark-matter particle 
with this coupling structure.
Unfortunately, we have already seen that our pseudoscalar $\Delta \tilde q^{(N)}$ 
coefficients, although significantly enhanced relative to their axial-vector counterparts,
are not large enough to overcome this degree of velocity suppression.
Thus we do not expect the operator ${\cal O}_{\chi q}^{\rm (PP)}$ to have much relevance
for direct-detection experiments. 
   
Let us now turn to the situation
in which $\chi$ primarily couples to SM particles 
through the operator ${\cal O}_{\chi q}^{\rm (SP)}$.  
In sharp contrast to the ${\cal O}_{\chi q}^{\rm (PP)}$ case discussed above, in this case 
only the quark bilinear gives rise to a velocity suppression in the cross-section for 
non-relativistic $\chi$/nucleon scattering.  This cross-section therefore scales like
\begin{equation}
   {\rm SP:}~~~~~~
  \sigma_{\mathrm{SD}}^{(\chi N)} ~\sim~ 
    \frac{c_q^2 \left[\Delta \tilde q^{(N)}\right]^2 \mu_{\chi N}^4 v^2}{m_N^2\Lambda^4}~.
\end{equation} 
Moreover, in this case we see that
dark-matter 
annihilation in the early universe is $p$-wave suppressed, since the initial 
state is CP-even, with quantum numbers $S=1$, $L=1$, and $J=0$. 
The annihilation cross-section in this case scales like
\begin{equation}
    {\rm SP:}~~~~~~
  \langle\sigma |v|\rangle ~\sim~ v_{\chi,\mathrm{fr}}^2\frac{c_q^2 m_\chi^2}{\Lambda^4}~,
\end{equation}
where $v_{\chi,\mathrm{fr}}$ denotes the average speed of $\chi$ and $\overline{\chi}$ 
at freeze-out.  Typically, $v_{\chi,\mathrm{fr}}^2 \sim 0.1$.  
Imposing, as before, the condition $\langle\sigma |v|\rangle \sim 1$~pb in order to 
ensure that $\Omega_\chi \approx \OmegaDM$, we find that 
\begin{equation}
    {\rm SP:}~~~~~~
  \sigma_{\mathrm{SD}}^{(\chi N)} ~\sim~ \left(1\mathrm{~pb}\right)\times 
     \frac{10 \left[\Delta \tilde q^{(N)}\right]^2 \mu_{\chi N}^4 v^2}{m_\chi^2 m_N^2}~.
\end{equation}
Since the velocity suppression $v^2 \sim \mathcal{O}(10^{-6})$ obtained in this case 
is far less severe than that obtained in Eq.~(\ref{eq:LPPv4Suppression}), we see that
only moderately large values for the $\Delta \tilde q^{(N)}$ coefficients are required in order to 
compensate for this velocity suppression and render the operator ${\cal O}_{\chi q}^{\rm (SP)}$
relevant for direct detection.  Moreover, as we have seen in Sect.~\ref{sec:NucleonCouplings},
these coefficients are indeed enhanced by the required amount.

We thus conclude that ${\cal O}_{\chi q}^{\rm (SP)}$, rather than ${\cal O}_{\chi q}^{\rm (PP)}$,
has greater prospects for being relevant to future direct-detection experiments.
As a result, we shall concentrate on ${\cal O}_{\chi q}^{\rm (SP)}$ in the remainder of this paper.


\section{Phenomenological consequences:  
Direct detection and related benchmarks \label{dirdet}}


We now turn to investigate the direct-detection prospects for a
dark-matter candidate in each of the three benchmark coupling scenarios 
defined in Sect.~\ref{CouplingScenarios}. 
In particular, we wish to determine the bounds imposed by 
existing direct-detection data on the corresponding suppression scale 
$M_{\rm I}$, $M_{\rm II}$, or $M_{\rm III}$ in each of these scenarios
as a function of the dark-matter mass $m_\chi$ and the coupling angle $\theta$,
and to assess the extent to which the next generation of direct-detection
experiments will be able to probe the remaining parameter space in each 
scenario.

In interpreting the results of such a direct-detection analysis, it is also useful 
to examine the relationship between the region of parameter space accessible by
direct-detection experiments in each of these coupling scenarios and regions 
of parameter space which are relevant for other aspects of dark-matter 
phenomenology.  For example, thermal freeze-out offers a natural mechanism for 
generating a relic abundance of the observed magnitude
for a massive dark-matter particle which can annihilate to SM particles. 
It is therefore interesting to 
examine whether successful thermal freeze-out can be realized within the
region of parameter space accessible to the next generation of 
direct-detection experiments for a dark-matter particle which annihilates 
primarily via ${\cal O}_{\chi q}^{\rm (SP)}$.  In addition, new-physics searches in a 
variety of channels at the LHC constrain the parameter space of operators which 
couple the dark and visible sectors.  It is therefore also interesting to examine 
the interplay between these constraints and those from direct-detection data.

The plan of this section is as follows. 
We begin by briefly reviewing the physics of direct detection 
and assessing the extent to which the next generation of direct-detection 
experiments will be capable of probing the parameter space of each of our 
benchmark coupling scenarios.  We then identify the regions of that parameter 
space which yield a thermal dark-matter relic abundance of the correct order, 
and discuss how LHC data serve to constrain 
that parameter space.  As we shall see, the magnitudes of the 
pseudoscalar $\Delta\tilde{q}^{(N)}$ coefficients
have a profound effect on the direct-detection 
phenomenology of a dark-matter particle which interacts with the visible 
sector primarily via the ${\cal O}_{\chi q}^{\rm (SP)}$ operators.

\subsection{Direct detection}

The principal physical quantity probed by direct-detection experiments is the 
total event rate $R$ for dark-matter scattering off the nuclei in the detector target.
For a generic dark-matter model, the expectation value for $R$ at any particular such 
experiment is obtained by integrating the differential rate $dR/dE_R$ over the range of 
recoil energies $E_R$ probed by that experiment, convolved with the 
appropriate detector-efficiency function ${\cal E}(E_R)$.  
This differential event rate (for reviews, see, \eg, Refs.~\cite{JungmanKamionkowskiGriest,LewinSmith})
is given by the general expression 
\begin{equation}
  \frac{dR}{dE_R} ~=~ \frac{N_T \rho_\chi^{\mathrm{loc}}}{m_\chi} 
    \int_{v> v_{\rm min}}^{\infty}v f(\vec{v})
    \left( \frac{d\sigma_{\chi T}}{d E_R} \right) d^3 v~, 
\label{eq:diffRate}
\end{equation}
where $N_T$ is the number of nuclei in the detector target, 
where $\rho_\chi^{\mathrm{loc}}$ is the local density of $\chi$ within the galactic halo,  
where $f(\vec{v})$ is the velocity distribution of dark-matter particles in the reference
frame of the detector, 
where $v\equiv |\vec v|$,
and where $d\sigma_{\chi T}/dE_R$ is the differential scattering 
cross-section.  The lower limit $v_{\rm min}$ on the integral over halo 
velocities corresponds to the kinematic threshold for non-relativistic scattering of
a dark-matter particle off one of the target nuclei.  

While substantial uncertainties exist concerning many of the aforementioned 
quantities which characterize the properties of the dark-matter halo, 
our focus in this paper is on the pseudoscalar nucleon 
coefficients $\Delta \tilde{q}^{(N)}$ and their implications for direct detection.  
We therefore adopt a set of standard benchmark assumptions about the dark-matter halo.
In particular, we take $\rho_\chi^{\mathrm{loc}} = 0.3\mathrm{~GeV\, cm}^{-3}$;
we take 
$f(\vec{v})$ to be Maxwellian, but truncated above 
the galactic escape velocity $v_{\mathrm{esc}}\approx 550$~km/s in the halo frame;
and we take 
$v_e = 232$~km/s as the speed of the Earth with respect to the 
dark-matter halo.
Moreover, we focus on the case in which $\chi$/nucleus scattering is purely 
elastic, for which ${v_{\rm min}=\sqrt{E_R m_T / 2 \mu_{\chi T}^2}}$, where $m_T$ 
denotes the mass of the target nucleus and where $\mu_{\chi T}$ is the reduced mass
of the $\chi/{\rm nucleus}$ system.

The differential cross-section for $\chi$/nucleus scattering is given by the general 
expression
\begin{equation}
  \frac{d\sigma_{\chi T}}{dE_R} ~=~ \frac{m_T}{2\pi v^2} 
     \left\langle|{\cal M}_{\chi T}|^2\right\rangle~,
     \label{eq:GeneraldSigmadER}
\end{equation}
where $\langle|{\cal M}_{\chi T}|^2\rangle$ is the corresponding squared $S$-matrix element,
averaged over initial spin states and summed over final spin states.  For the 
scalar-pseudoscalar interaction we are considering here, we recall Eq.~(\ref{limits})
to find that this matrix element in the non-relativistic limit 
takes the form
\begin{align}
{\cal M}_{\chi T} &~=~ 
   \sum_{N=n,p} \frac{g_{\chi N}}{\Lambda^2} \langle\chi_f |\overline{\chi}\chi |\chi_i\rangle
   \langle T_f | \overline{N}\gamma^5 N | T_i \rangle \nonumber \\ 
   &~\approx ~
   \frac{4 m_{\chi} m_T}{\Lambda^2} (\xi_{\chi}^{s'})^\dagger  \xi_{\chi}^{s} 
   \sum_{N=n,p} \frac{g_{\chi N}}{m_N} \langle T_f | \vec q \cdot \vec {S}_{N} | T_i \rangle~,
   \label{eq:MatElChiNucPS}
\end{align}
where $\langle T_f|\vec S_{N}|T_i\rangle$ denotes 
the matrix element for the nucleon-spin operator within the target nucleus 
and where $\vec q$ is the momentum transferred to the nucleus. 
Note that the $m_T/m_N$ factor in Eq.~(\ref{eq:MatElChiNucPS})
arises due the difference in normalization 
between the constituent nucleons and the bound-state nucleus,
where we have retained the relativistic normalization in both cases.
Proceeding by analogy with the axial-vector case~\cite{GoodmanWitten:1985}, we
invoke the Wigner-Eckart theorem in order to make the replacement
\begin{equation}
\langle T_f | \vec {S}_N | T_i \rangle ~\rightarrow~  
  \frac{\langle S_N \rangle}{J_T}\, \langle T_f | \vec{J}_{T} | T_i \rangle~
\end{equation}
in Eq.~(\ref{eq:MatElChiNucPS}), where 
$\langle S_N \rangle / J_T = \langle T_f | S_N | T_i \rangle / J_T $ 
again represents the fraction of the total nuclear spin carried by the nucleon $N$.
In the approximation that $m_p \approx m_n$, this yields 
\begin{eqnarray}
  {\cal M}_{\chi T} &=& \frac{4 m_\chi m_T}{J_T \Lambda^2 m_N}
  \biggl( g_{\chi p} \langle S_p \rangle + g_{\chi n} \langle S_n \rangle\biggr)
  (\xi_{\chi}^{s'})^\dagger \xi_{\chi}^{s}\nonumber\\
  &&~~~~~ \times ~ \langle T_f|\vec q \cdot \vec J_{T}|T_i\rangle~.
\end{eqnarray}
The spin-averaged squared matrix element is therefore
\begin{align}
\left \langle |{\cal M}_{\chi T}|^2\right\rangle &= 
  \frac{16 m_{\chi}^2 m_{T }^2}{J_{T}^2(2J_T+1)m_N^2 \Lambda^4}
  \biggl( g_{\chi p} \langle S_p \rangle + g_{\chi n} \langle S_n \rangle\biggr)^2 \nonumber \\
  &~~~~~~~ \times \sum_{T_i,T_f}
  \langle T_f | \vec q \cdot \vec J_{T} | T_i \rangle \langle T_i | \vec q \cdot \vec{J}_{T} | T_f \rangle \nonumber \\
    ~&=\frac{16 m_{\chi}^2 m_T^2 |\vec{q}|^2 }{3m_N^2}
      \frac{J_T{+}1}{J_T}
  \left(\frac{g_{\chi p}}{\Lambda^2} \langle S_p \rangle + \frac{g_{\chi n}}{\Lambda^2}
  \langle S_n \rangle\right)^2.
\end{align}
Substituting this result into Eq.~(\ref{eq:GeneraldSigmadER}) and dividing by 
$16 m_{\chi}^2 m_{T }^2$ in order to account for the difference between relativistic 
and non-relativistic normalization conventions for the $\chi$ and nucleus states, we 
arrive at our final expression for the differential cross-section for $\chi$/nucleus scattering:
\begin{align}
\frac{ \partial \sigma_{\chi T}^{\rm (SP)}}{\partial E_R} ~&=~ 
  \frac{m_T^2 E_R}{3 \pi v^2 m_N^2}\frac{J_T + 1}{J_T} 
  \nonumber \\ & ~~~~~ \times
  \left(\frac{g_{\chi p}}{\Lambda^2} \langle S_p \rangle + 
  \frac{g_{\chi n}}{\Lambda^2} \langle S_n \rangle\right)^2
  \widetilde{F}^2(E_R)~,~~
  \label{eq:diffXsecPS}
\end{align}
where $\widetilde{F}^2(E_R)$ is a nuclear form factor.  Note that we have explicitly 
distinguished this form factor from the usual form factor $F^2(E_R) = S(E_R)/S(0)$ 
associated with spin-dependent scattering via an axial-vector interaction.  Indeed, in 
the axial-vector case, the scattering cross-section depends on the projection of $\vec{S}_N$
along the direction of the spin vector $\vec{S}_\chi$ of the dark-matter particle.  
By contrast, in the scalar-pseudoscalar case, the corresponding cross-section depends on the 
projection of $\vec{S}_N$ along the direction of the momentum 
transfer~\cite{FitzpatrickMathematica}. 

A wealth of data from direct-detection experiments already significantly 
constrains the set of possible interactions between dark-matter particles and 
atomic nuclei, and several additional experiments are poised to probe even more 
deeply over the coming decade into the parameter space of allowed couplings 
between the dark and visible sectors.  For each of our three benchmark coupling scenarios 
for scalar-pseudoscalar interactions, the relevant parameter space comprises   
$m_\chi$, $\theta$, and the corresponding suppression scale $M_{\rm I}$, $M_{\rm II}$, 
or $M_{\rm III}$.  The first of these parameters enters the expected event rate for 
a given detector in a complicated way through the scattering kinematics, while 
the second and third enter through the ratios 
$g_{\chi p}/\Lambda^2$ and $g_{\chi n}/\Lambda^2$ in Eq.~(\ref{eq:diffXsecPS}), 
as discussed in Sect.~\ref{CouplingScenarios}.

Since the $\chi$/nucleus interactions which follow from ${\cal O}_{\chi q}^{\rm (SP)}$ 
involve the nuclear spin $\vec{S}_N$, the relevant constraints on these parameters 
are those which pertain to spin-dependent scattering.  Several 
direct-detection experiments 
already provide comparable, stringent limits on spin-dependent 
scattering~\cite{COUPP4Limits,SIMPLE,PICASSO}.  Moreover, the next generation of these 
experiments, including COUPP-60 and PICO-250L, are projected to significantly extend   
the reach of these experiments in the near future~\cite{COUPPAspenTalk}.
In this paper, our primary aim is to investigate the sensitivity of these latter experiments 
to scalar-pseudoscalar interactions between dark-matter particles and atomic nuclei.  We 
therefore focus on the results from COUPP-4, for which the experimental setup and analysis 
parallel those for COUPP-60 and PICO-250L, when discussing existing limits on 
spin-dependent scattering.  These limits are typically expressed as bounds on 
the spin-dependent dark-matter/proton scattering cross-section 
$\sigma_{\chi p}^{\rm (AA)}$ for a dark-matter 
particle whose interactions with nuclei are primarily due to the axial-vector operators 
${\cal O}_{\chi q}^{\rm (AA)}$.  This cross-section may be parametrized as   
\begin{equation}
  \sigma_{\chi p}^{\rm (AA)} ~=~ \frac{3 a_{\chi p}^2\mu_{\chi p}^2}{\pi \Lambda^4}~,
\end{equation} 
where $\mu_{\chi p}$ is the reduced mass of the $\chi$/proton system and where
\begin{equation} 
  a_{\chi N} ~\equiv ~ \sum_{q=u,d,s} c_q^{\rm (AA)}\Delta q^{(N)}
\end{equation}
are the axial-vector analogues of the $\chi$/nucleon couplings $g_{\chi N}$ given in 
Eq.~(\ref{eq:gchiNgeneral}).
Note that because we take $\chi$ to be a Dirac fermion, this expression differs by a 
factor of $4$ from the standard expression for a Majorana fermion.
The differential cross-section for $\chi$/nucleus scattering for such an interaction,
expressed in terms of $\sigma_{\chi p}^{\rm (AA)}$, is given by   
\beqn
\frac{\partial \sigma_{\chi T}^{\rm (AA)}}{\partial E_R}  &=&
  \frac{2 \sigma_{\chi p}^{\rm (AA)} m_T}{3 \mu_{p}^2 v^2}\frac{J_T + 1}{J_T}\nonumber\\
  &&~~~ \times ~ \left(\langle S_p \rangle + \frac{a_{\chi n}}{a_{\chi p}}\langle S_n \rangle \right)^2 
  F^2(E_R)~.~~~~~~~~     
\label{eq:diffXsec}
\eeqn
It is therefore straightforward to convert the limits on $\sigma_{\chi p}^{\rm (AA)}$ into 
limits on the expected event rate for dark-matter scattering off nuclei within the 
detector volume.  The latter limits are model-independent and applicable to 
any interaction between dark-matter and atomic nuclei, including the scalar-pseudoscalar 
interactions which are the focus of this paper.

The bounds implied by COUPP-4 data on the parameter space of each of our 
three coupling scenarios, along with the projected reach into that parameter 
space for both COUPP-60 and PICO-250L, will be discussed in the next section.
These bounds and sensitivities will be expressed as 
contours in $(m_\chi,M_\ast)$ space
for each scenario and for several benchmark values of $\theta$, where $M_\ast$ denotes 
the corresponding suppression scale $M_{\rm I}$, $M_{\rm II}$, or $M_{\rm III}$. 
In evaluating these contours, we will make use of the DMFormFactor 
package~\cite{FitzpatrickMathematica}.  We will also include bands indicating the 
uncertainties in these contours which arise as a result of the uncertainties in the 
nucleon couplings $g_{\chi N}$ discussed in Sect.~\ref{sec:NucleonCouplings}.

\subsection{Relic abundance\label{sec:Annihilation}}

Thermal freeze-out is a natural mechanism through which a sizable relic 
abundance can be generated for a massive particle with suppressed couplings 
to SM states.  It is therefore useful to  
identify the regions of parameter space within which the relic abundance of a 
dark-matter particle which annihilates via the ${\cal O}_{\chi q}^{\rm (SP)}$ operator
reproduces the observed dark-matter relic abundance $\OmegaDM \approx 0.26$~\cite{Planck}.
In this section, we briefly summarize the relic-abundance calculation for an
interaction of this sort.  Note that we take $\chi$ to be a Dirac fermion throughout 
and make use of the general formalism in Ref.~\cite{EdsjoGondolo} for multi-particle
freeze-out dynamics in order to evaluate the total relic abundance of $\chi$ and its conjugate
$\overline{\chi}$, which in this case represent distinct degrees of freedom.

The evolution of the total number density $Y \equiv Y_\chi + Y_{\bar{\chi}}$ of 
particles which contribute to the dark-matter abundance at late times due to thermal 
freeze-out in this scenario can be described by the single differential 
equation
\begin{equation}
  \frac{dY}{dt} ~=~ - s \, \langle\sigma|v|\rangle \,
     \big[Y^2 - (\Yeq)^2\big]~,
  \label{eq:ChiDiffEqTime}
\end{equation}
where $\Yeq$ is the value which $Y$ would have were $\chi$ and $\overline{\chi}$ 
in thermal equilibrium at time $t$; where $s= 2\pi^2 g_{\ast s}(T)T^3 / 45$ is
the entropy density of the universe, expressed here in terms of the temperature $T$
of the thermal bath at time $t$ and the number of effectively massless degrees of 
freedom $g_*(T)$ at that temperature $T$; and where $\langle\sigma|v|\rangle$ is the 
thermally-averaged total cross-section for dark-matter annihilation.
The total present-day dark-matter-abundance contribution from $\chi$ and 
$\overline{\chi}$ due to thermal freeze-out is related to the present-day value
$Y_{\mathrm{now}}$ of $Y$ by 
\begin{equation}
  \Omega_\chi ~\equiv~ \frac{\rho_\chi}{\rhocrit} ~=~
  \frac{\snow \, m_\chi \, Y_{\mathrm{now}}}{\rhocrit}~,
\end{equation}
where $\snow \approx 2.22\times 10^{-38}\mathrm{~GeV}^3$
and $\rhocrit \approx 4.18 \times 10^{-47}\mathrm{~GeV}^4$ are the present-day
entropy density and present-day critical energy density of the universe,
respectively.

In the case in which $\chi$ and $\overline{\chi}$
annihilate primarily to SM quarks via ${\cal O}_{\chi q}^{\rm (SP)}$, 
we find that the thermally-averaged annihilation cross-section for 
processes of the form $\overline{\chi}\chi\rightarrow \overline{q}q$ is
given by
\begin{equation}
  \langle\sigma|v|\rangle ~=~
    \frac{3x}{256\pi m_\chi^5 K_2(x)}\sum_q \frac{c_q^2}{\Lambda^4} {\cal I}_q(x)~,
  \label{eq:SigmavFinal}
\end{equation}
where we have defined
\begin{equation}
  {\cal I}_q(x) ~\equiv~ \int_{4m_\chi^2}^\infty ds
    \sqrt{s(s-4m_\chi^2)^3 (s-4m_q^2)}
    \, K_1\left(\frac{x\sqrt{s}}{m_\chi}\right)~.
  \label{eq:Iq}
\end{equation}
In these expressions, $x \equiv m_\chi / T$, 
$s = (p_\chi + p_{\bar{\chi}})^2$ is the usual Mandelstam
variable ({\it not}\/ the entropy density of the universe, and {\it not}\/ the 
strange quark either), and 
$K_1(x)$ and $K_2(x)$ denote the modified Bessel functions of
the second kind of degree one and two, respectively.  

In the next section we will 
display contours corresponding to the condition 
$\Omega_\chi = \OmegaDM$, 
as well as 
contours of $\langle \sigma|v|\rangle$.  In accord with 
expectation, we will find that a relic abundance of the correct order 
is obtained for 
$\langle \sigma|v|\rangle \approx 1$~pb.  In interpreting these results,
it should be noted that $\Omega_\chi$ depends on $m_\chi$ in
the usual manner, whereas this quantity depends on $\theta$ and the 
corresponding suppression scale $M_{\rm I}$, $M_{\rm II}$, or $M_{\rm III}$ 
in each of our coupling scenarios through the
ratio ${c_q^2/\Lambda^4}$ in Eq.~(\ref{eq:SigmavFinal}).
Generally speaking, $\Omega_\chi\propto\langle \sigma|v|\rangle^{-1}$
for thermal freeze-out, and therefore a higher suppression 
scale corresponds to a smaller $\langle\sigma|v|\rangle$ and a larger 
relic abundance.

\subsection{Collider constraints\label{sec:ColliderLimits}}

Colliders offer a complementary way of probing the couplings between dark-sector
and visible-sector fields.  In particular, the effective operators given in 
Eq.~(\ref{operator}) generically contribute to the event rate for 
processes of the form $pp\rightarrow X + \displaystyle{\not}E_T$ at the LHC ---
\ie, so-called ``mono-anything'' processes --- 
where $X$ denotes a single SM particle such as a photon (the monophoton channel), 
an electroweak gauge boson, or even a ``particle'' such 
as a hadronic jet (the monojet channel). 
While the results depend on the particular operator and the relative values of the 
coupling coefficients (see, \eg, Ref.~\cite{ColliderDMBlock}),
the most stringent constraints on such operators are typically 
those derived from limits on monojet production at ATLAS~\cite{ATLASMonojet,ATLASMonojet8TeV} 
and CMS~\cite{CMSMonojet,CMSMonojet8TeV} and from limits on the production of a 
hadronically decaying $W^\pm$ or $Z$ boson at ATLAS~\cite{monoWZ}.  We will henceforth 
focus on these channels, but we also note that a combined 
analysis~\cite{MonoCombinedCheung,MonoEverything} involving all relevant
$pp\rightarrow X+ \displaystyle{\not}E_T$ processes would lead 
to a slight enhancement of the bounds from these two leading channels individually.  
Moreover, we also note that searches in the mono-$b$ and $t\overline{t} + \displaystyle{\not}E_T$ channels 
can potentially
supersede these limits for models in which the couplings between the 
dark matter and the third-generation quarks are
enhanced~\cite{monob}, as is the case in our Scenario~II.

We now proceed to derive a set of rough limits on the corresponding suppression 
scale $M_{\rm I}$, $M_{\rm II}$, or $M_{\rm III}$ associated with each 
of our benchmark coupling scenarios.  We derive these limits under the 
assumption that a contact-operator description of the interactions between 
$\chi$ and the SM quarks remains valid up to the center-of-mass-energy scale 
$\sqrt{s}\approx 8$~TeV of the LHC.~  We then return to discuss how these results 
are altered in cases in which the contact-operator description is valid at 
scales relevant for direct detection, but breaks down at scales well below 
$\sqrt{s}$.     
  
We begin by noting that the monojet~\cite{CMSMonojet8TeV} and 
mono-$W/Z$~\cite{monoWZ} analyses which correspond to the most stringent 
current limits on dark-matter production at the LHC are effectively 
counting experiments which serve to constrain the total cross-section 
for the corresponding production processes.  A lower limit 
$M_\ast > M_{\mathrm{min}}$ on the heavy mass scale $M_\ast$ defined
for the operator D3 in the standard operator-classification scheme of 
Ref.~\cite{TimLimits} [which corresponds to our scalar-pseudoscalar 
operator ${\cal O}_{\chi q}^{\rm (SP)}$] from either of these analyses corresponds to a limit 
$M_{\rm II} > M_{\mathrm{min}}/2^{1/6}$ in our Scenario~II with 
$\theta = \pi/4$.  We also note that the production cross-section for 
each process scales like 
$\sigma_{\rm I}(m_\chi,M_{\rm I},\theta) \propto M_{\rm I}^{-4}$ 
in Scenario~I, whereas it scales like 
$\sigma_{\rm II,III}(m_\chi,M_{\rm II,III},\theta) \propto M_{\rm II,III}^{-6}$
in Scenarios~II and~III.~  It therefore follows that bounds on 
$M_{\rm I}$ can be derived from the bounds on $M_{\mathrm{min}}$ quoted
in Refs.~\cite{CMSMonojet8TeV} and~\cite{monoWZ} and the ratio of 
the corresponding production cross-sections for the same 
$m_\chi$ and the same fiducial value of $M_{\rm I}$.  
In this analysis, we choose $1$~TeV as our fiducial mass scale.
We therefore have   
\begin{equation}
  \frac{M_{\rm I}}{\mathrm{GeV}} ~\gtrsim ~ 
  \left[\frac{\sigma_{\rm I}(m_\chi,1\mathrm{\,TeV},\theta)}
  {2\,\sigma_{\rm II}(m_\chi,1\mathrm{\,TeV},\pi/4)}\right]^{1/4}
  \left(\frac{M_{\mathrm{min}}}{10\mathrm{\,GeV}}\right)^{3/2}~.
\end{equation} 
Likewise, lower limits on $M_{\rm II}$ and $M_{\rm III}$ 
may be derived using the relation
\begin{equation}
  M_{\rm II,III} ~\gtrsim ~ \left[\frac{\sigma_{\mathrm{II,III}}
  (m_\chi,1\mathrm{\,TeV},\theta)}
  {2\,\sigma_{\rm II}(m_\chi,1\mathrm{\,TeV},\pi/4)}\right]^{1/6}
  M_{\mathrm{min}}~.
\end{equation} 

Constraint contours corresponding to the limits on contact-operator interactions from these
most recent monojet and mono-$W/Z$ analyses will be discussed in the next section
for each of our three coupling scenarios.  The relevant cross-sections in each case 
are evaluated at parton level using the MadGraph/MadEvent package~\cite{MadGraph} 
(with the CTEQ6L1 PDF set~\cite{CTEQPDFs}) including the contribution from processes 
involving $b$ quarks in the initial state.
The event-selection criteria we employ in estimating these limits are modelled on those
described in Ref.~\cite{CMSMonojet8TeV} for the mono-jet channel
and Ref.~\cite{monoWZ} for the mono-$W/Z$ channel,
and we have verified that minor alterations in these cuts do not have significant effects on our results.

As mentioned above,
it is important to note that constraints derived in this manner are valid 
only in the regime in which interactions between dark-matter particles and 
SM quarks can legitimately be modeled as contact operators at energies 
comparable to $\sqrt{s}$.  In other words, they are 
valid for processes in which the mass $m_\phi$ of the particle $\phi$ 
which mediates the interaction is much larger than the
momentum transfer to the dark-matter system.  By contrast, for lower mediator 
masses $m_\phi \lesssim 1~\tev$, these limits are no longer applicable ---
even for $m_\phi > m_\chi$.  Constraints on interactions between 
dark-matter particles and SM quarks can still be derived from LHC data for theories 
in which $m_\phi \lesssim 1~\tev$; however, such constraints 
are highly model-dependent, sensitive to the full structure of the dark sector, and 
frequently weaker than the na\"\i ve limits one would obtain for these same channels 
in the contact-operator regime~\cite{LowMassMediatorsLHC}. 

On the other hand, while the contact-operator approximation becomes unreliable
from the perspective of collider phenomenology for $m_\phi \lesssim 1~\tev$, it
remains valid for direct-detection phenomenology down to far lower values of $m_\phi$.
Indeed, interactions involving light mediators can still be reliably modeled as 
contact interactions at energies relevant for direct detection, provided that 
$m_\phi \gsim 1~\gev$.  Moreover, the relic-density calculation in 
Sect.~\ref{sec:Annihilation} also remains qualitatively unaltered in the presence
of a light mediator down to the kinematic threshold $m_\phi = m_\chi$.  Below this
threshold, annihilation into pairs of on-shell mediators becomes kinematically 
accessible. 
Moreover, below this threshold, 
the behavior of the thermally-averaged annihilation cross-section 
transitions from $\langle\sigma |v| \rangle \propto m_\chi^2 / m_\phi^4$ to
$\langle\sigma |v|\rangle \propto 1 / m_\chi^2$  
because $m_\chi$ is always the dominant energy scale entering into the propagators 
for all diagrams contributing to this annihilation cross-section.
Above this kinematic threshold, by contrast, we find that the 
correct relic density can be obtained for perturbative couplings between 
$\phi$ and both the dark-sector and visible-sector fermions in our theory, 
provided that $m_\chi \lesssim {\cal O}(10~\tev)$.

In light of these considerations, we emphasize that the monojet and mono-$W/Z$ limits
we have discussed here
should not be interpreted as exclusion bounds,
but rather as relations which indicate the regions within which LHC data can be interpreted
as requiring that 
the mediator particle(s) $\phi$ 
not be particularly heavy.  Indeed, the suppression scale 
$M_{\rm I}$, $M_{\rm II}$, or $M_{\rm III}$ in each of our three coupling scenarios can
still be large even if $m_\phi$ is light, provided the coupling between $\phi$ and
either $\chi$ or the SM quarks is small.


\section{Results}


\begin{figure*}[t]
\begin{center}
  \epsfxsize 2.25 truein \epsfbox {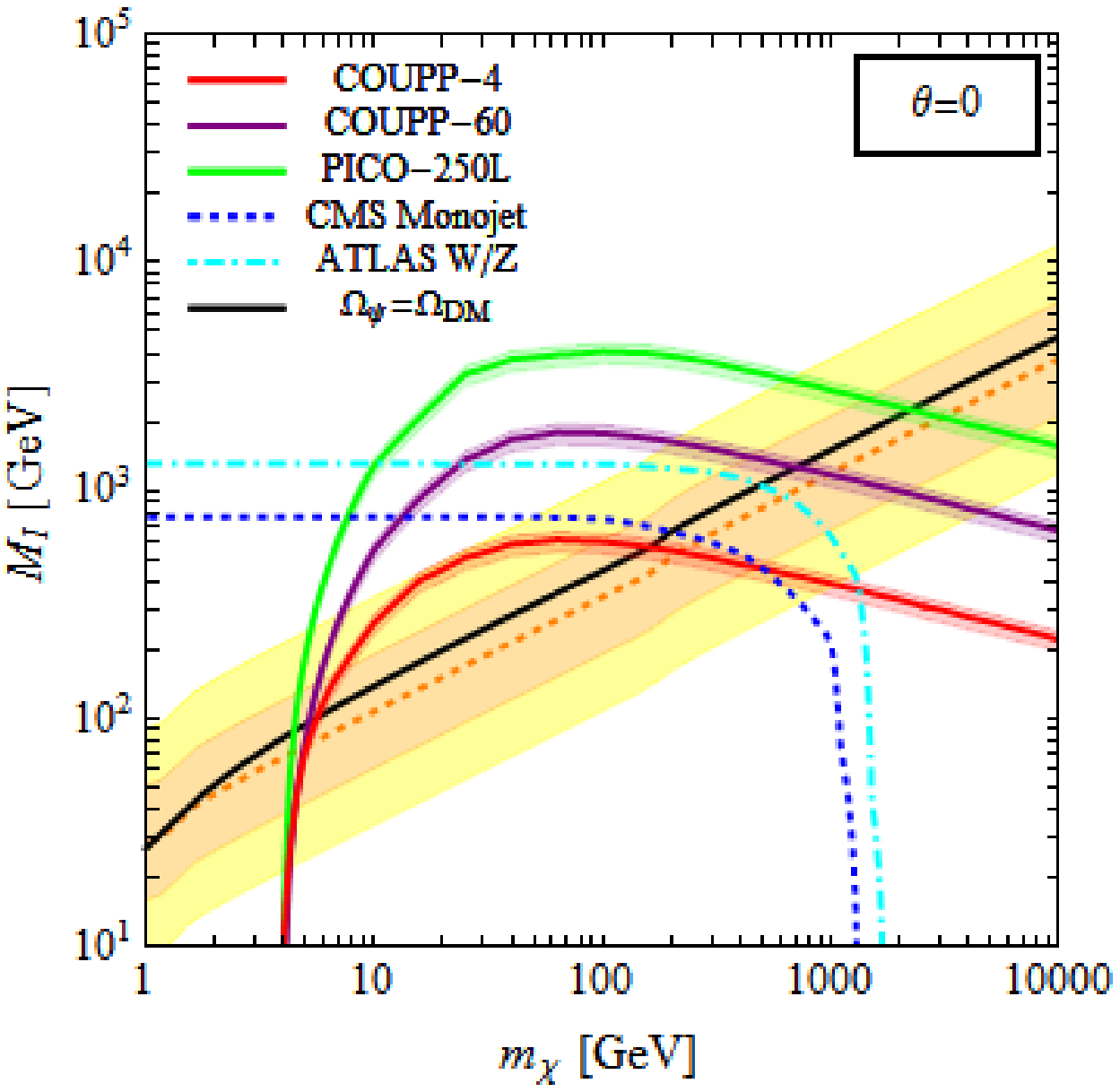}~
   ~~~
  \epsfxsize 2.25 truein \epsfbox {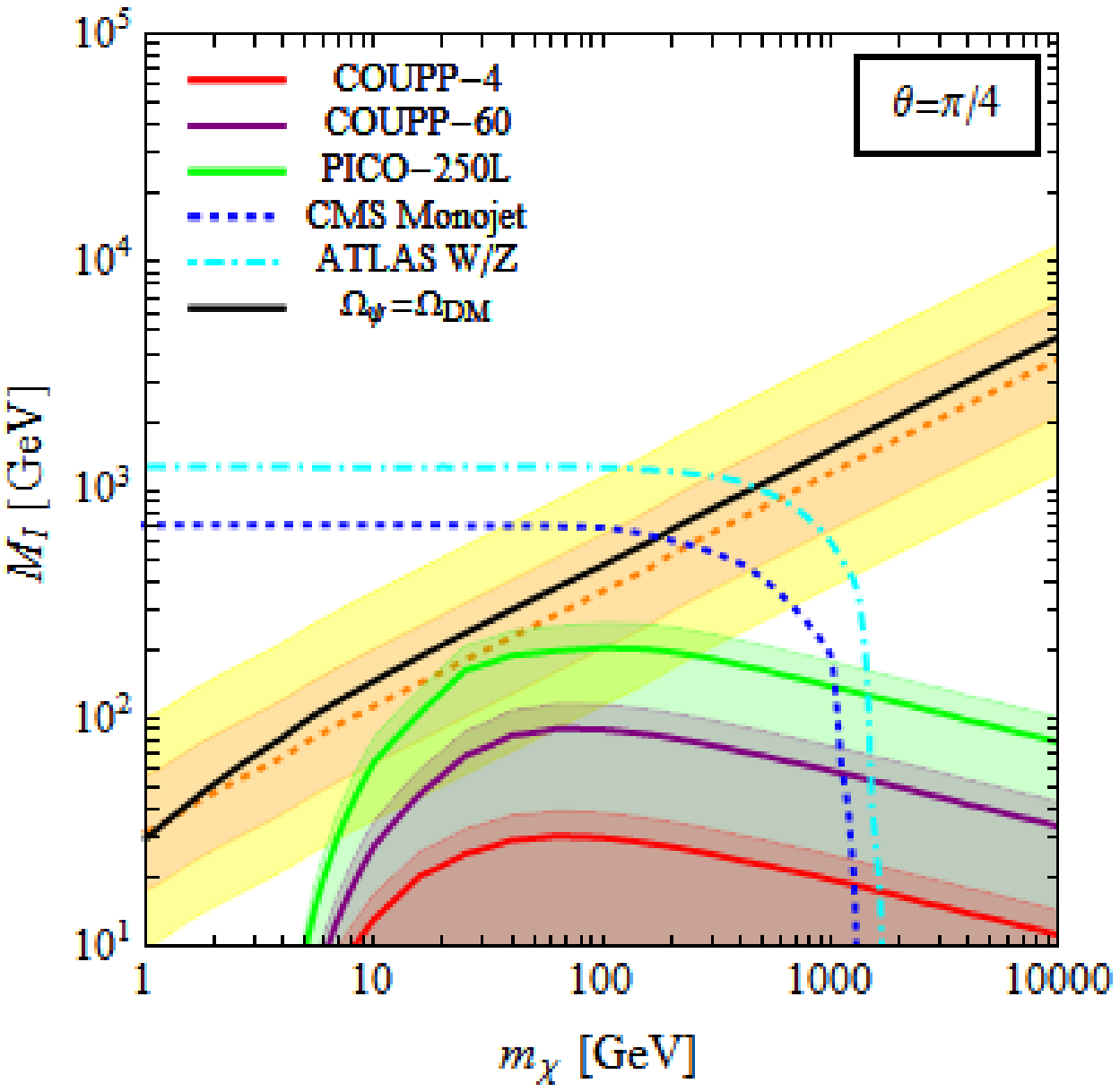}~
   ~~~
  \epsfxsize 2.25 truein \epsfbox {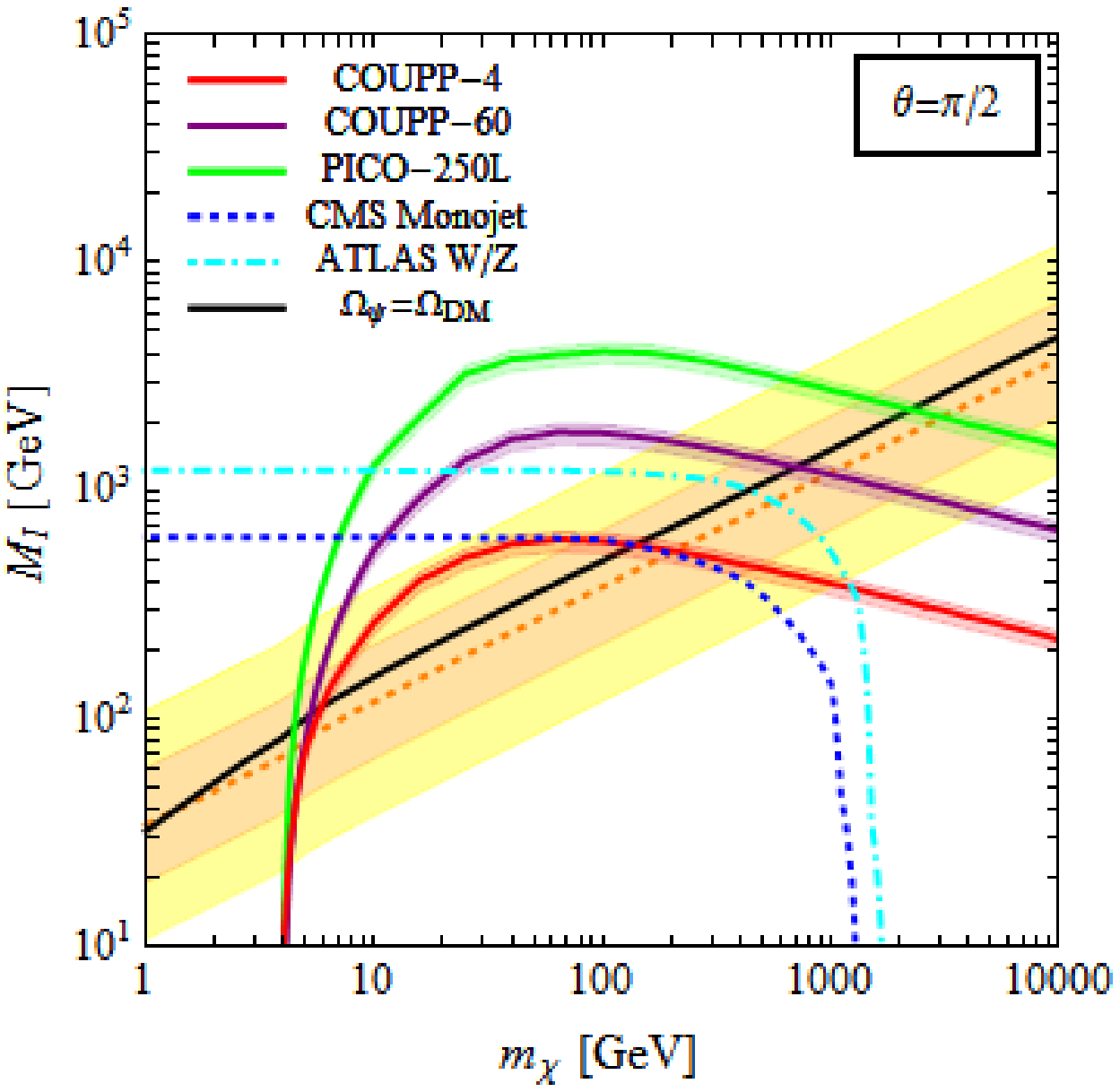}
\end{center}
\begin{center}
  \epsfxsize 2.25 truein \epsfbox {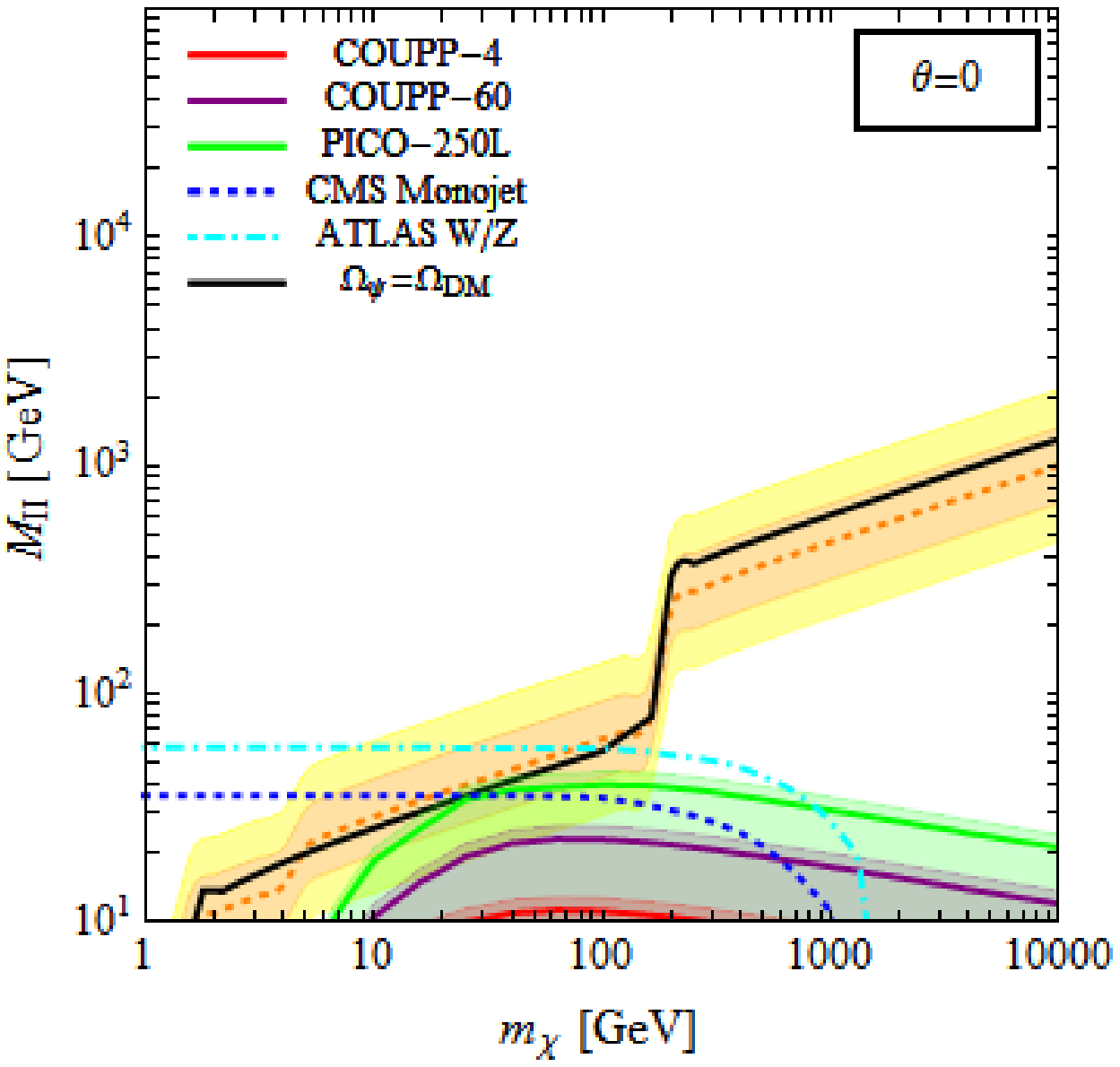}~
   ~~~
  \epsfxsize 2.25 truein \epsfbox {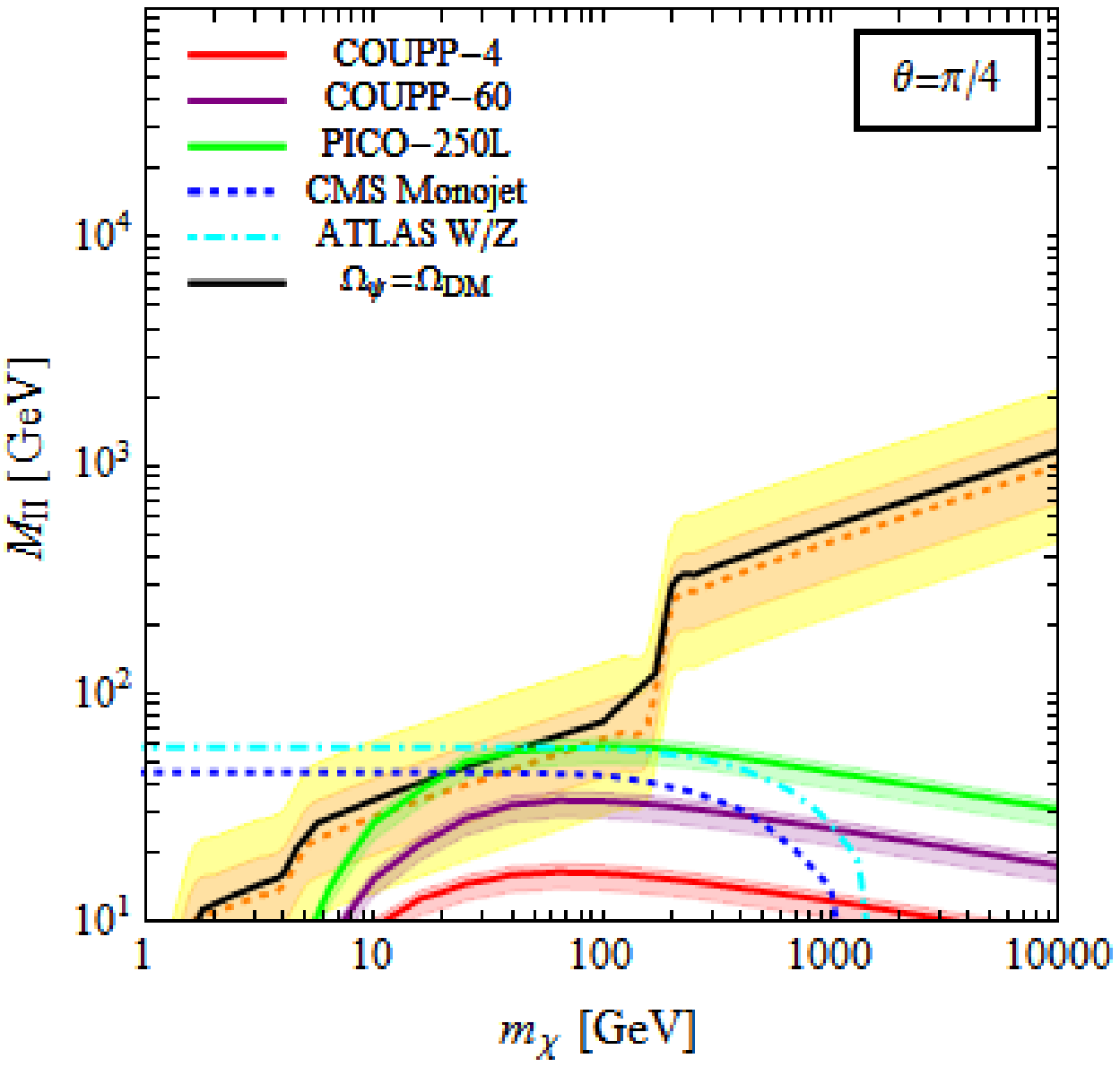}~
   ~~~
  \epsfxsize 2.25 truein \epsfbox {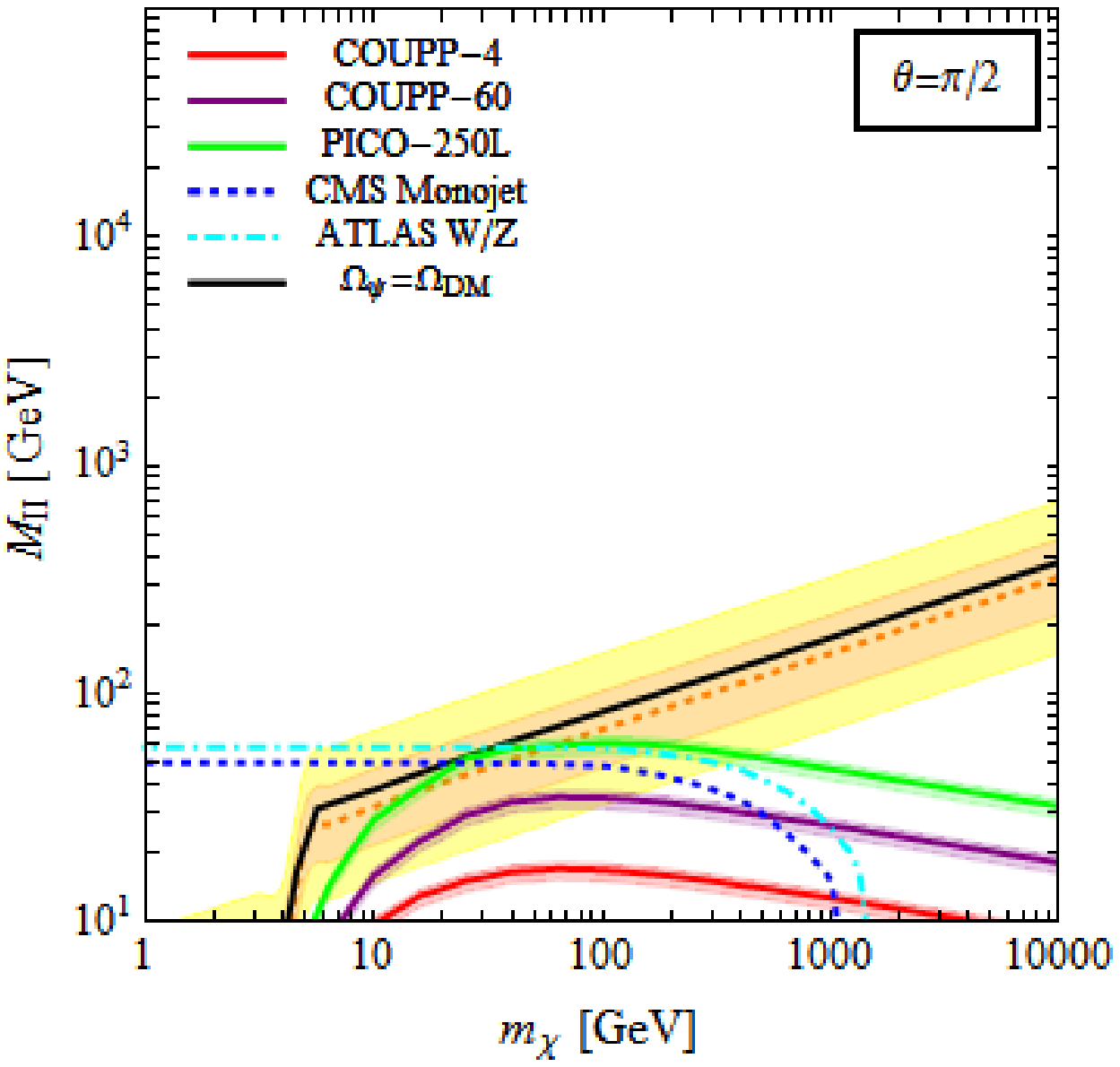}
\end{center}
\begin{center}
  \epsfxsize 2.25 truein \epsfbox {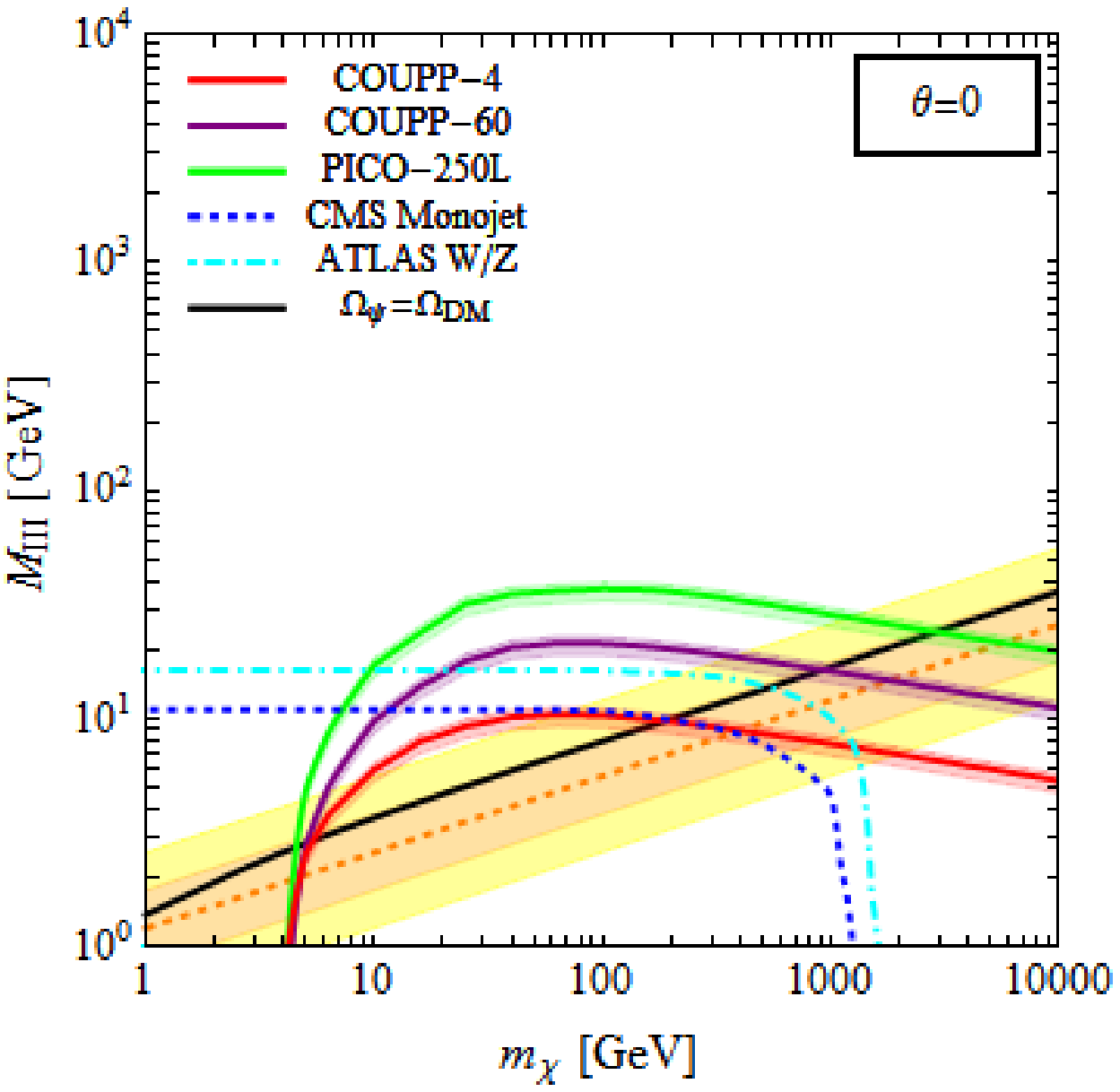}~
   ~~~
  \epsfxsize 2.25 truein \epsfbox {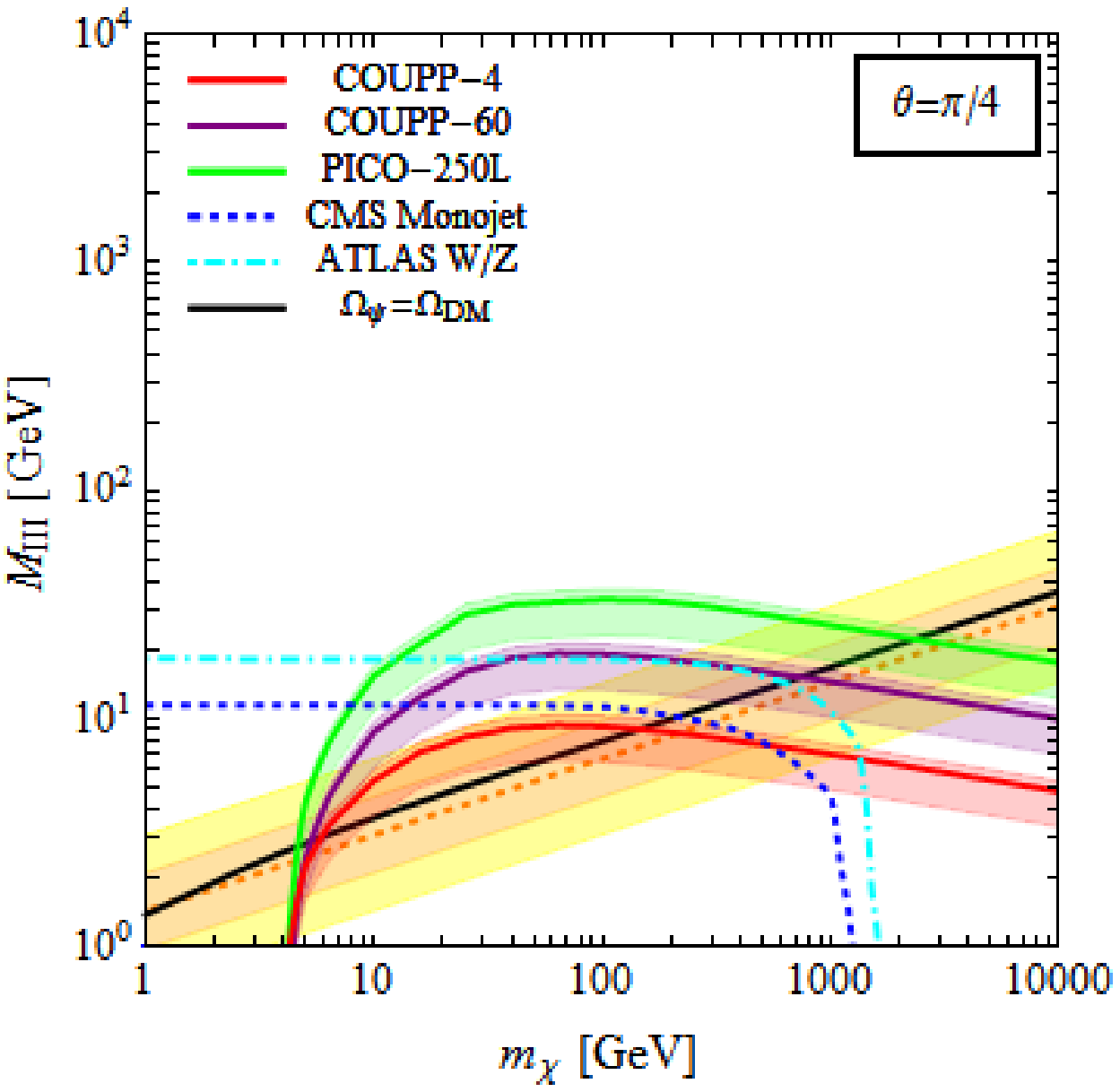}~
   ~~~
  \epsfxsize 2.25 truein \epsfbox {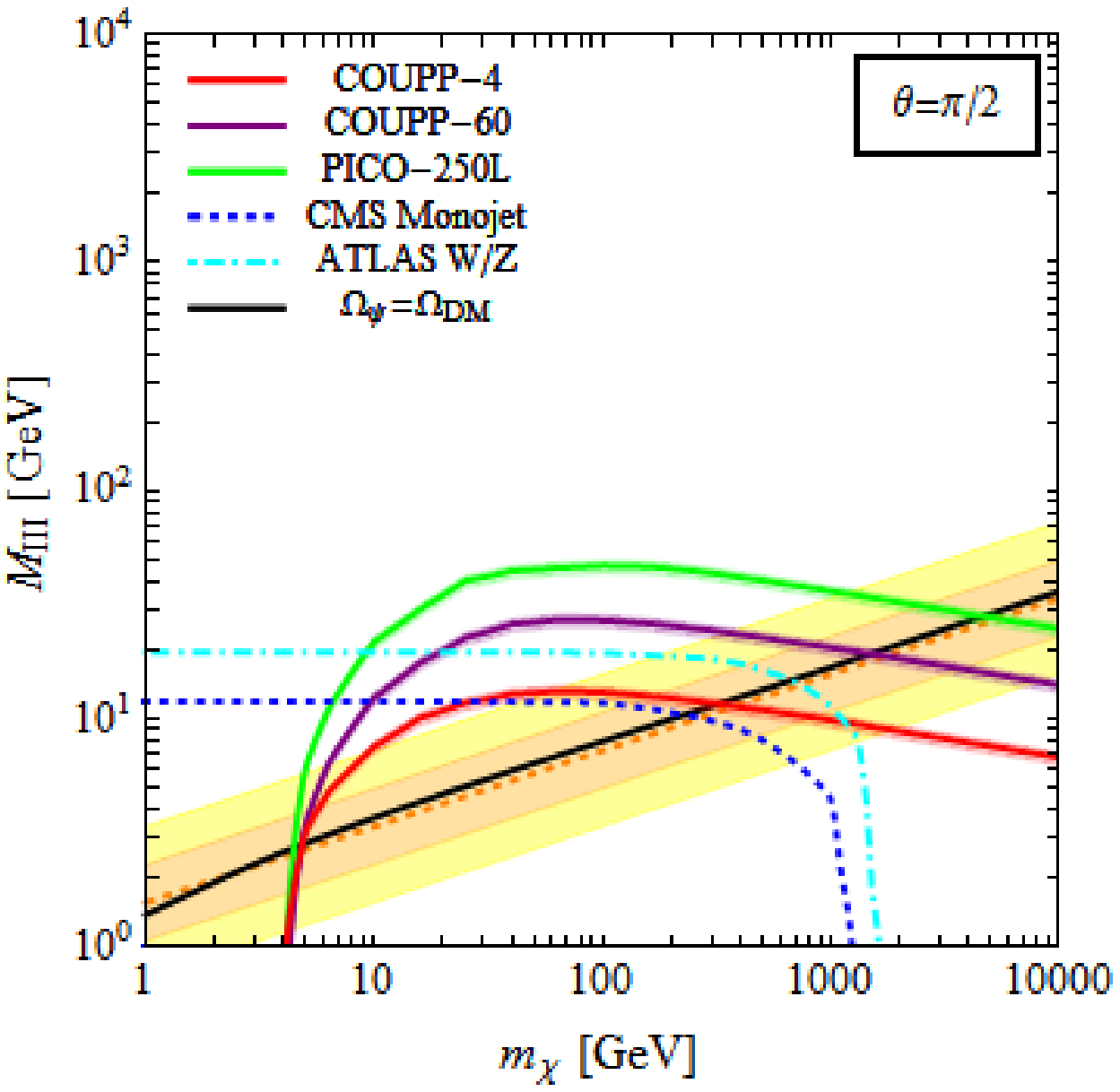}
\end{center}
\vskip -0.2 truein
\caption{
Experimental reach of direct-detection experiments, 
assuming pseudoscalar
interactions with the benchmark coupling structures of our Scenario~I (top row), 
Scenario~II (center row), and Scenario~III (bottom row), with $\theta=0$ (left column),
$\theta=\pi/4$ (center column), and $\theta=\pi/2$ (right column) in each case.
These coupling structures are discussed in Sect.~\protect\ref{CouplingScenarios},
and each panel is plotted as a function of the 
dark-matter mass $m_\chi$ and the mass scale $M_{\rm I}$, $M_{\rm II}$, or $M_{\rm III}$
associated with the corresponding scenario.
Within each panel,
the red curve indicates the upper limit of the region already excluded by COUPP-4 data, 
while the purple and green curves respectively indicate the projected discovery 
reaches of the COUPP-60 and PICO-250L experiments.  
The thickness of each curve 
indicates the uncertainty associated with the corresponding experimental reach, 
as discussed and calculated in the text;
note that in some cases these uncertainties are sufficiently large as to cause
these ``lines'' to become entire red-, purple-, and blue-shaded regions.
For guidance, we have also indicated the contour (black line) along which
$\Omega_\chi = \OmegaDM \approx 0.26$,
while the orange dashed line corresponds to
a thermally-averaged dark-matter annihilation cross-section 
$\langle \sigma |v|\rangle = 1$~pb. 
The pale peach- and yellow-colored bands correspond to the regions within which 
$0.1\mathrm{~pb} \leq \langle \sigma |v|\rangle \leq 10\mathrm{~pb}$ and 
$0.01\mathrm{~pb} \leq \langle \sigma |v|\rangle \leq 100\mathrm{~pb}$, respectively.
Finally, the blue dashed curve and cyan dot-dashed curve 
respectively indicate the lower limits on the appropriate mass scale $M_{\rm I,II,III}$ from monojet and 
mono-$W/Z$ searches at the LHC which would apply in the case of a heavy mediator.
We see from these plots that there are many situations in which upcoming 
direct-detection 
experiments can easily reach into the range of greatest interest for thermally-produced dark matter
and its possible collider signatures  --- even when only pseudoscalar interactions between
dark matter and Standard-Model quarks are assumed.}
\label{fig:Abundances}
\end{figure*}

In the previous section, 
we outlined the physics that determines the reach of various direct-detection experiments,
assuming only pseudoscalar interactions between dark matter and Standard-Model quarks.
We also outlined the physics that determines the cosmological dark-matter abundances after 
freeze-out,
and summarized the physics that determines
the reach of monojet and mono-$W/Z$ searches at the LHC.~
As we saw in Sect.~\ref{dirdet}, all of these calculations depend to varying degrees
on the particular flavor coupling structure assumed
(\ie, whether we are operating within Scenario~I, Scenario~II, or Scenario~III),
and on the particular value of $\theta$ in each case.

The results of these analyses are shown in Fig.~\ref{fig:Abundances}.
The reaches of the current and future direct-detection experiments
considered in this study
are shown in red, purple, and blue (along with their associated uncertainties);
for the COUPP-60 experiment we have assumed an exposure of $10^5$~kg$\,$d while
for the PICO-250L experiment we have assumed three years of 
running with a $500$~kg fiducial mass~\cite{COUPPAspenTalk}. 
Likewise, the black contour in each case corresponds 
to the condition $\Omega_\chi = \Omega_{\rm DM}$, which one would 
na\"\i vely expect to occur for 
$\langle \sigma |v|\rangle \approx 1$~pb.
The orange dashed curve, by contrast, explicitly indicates the
points for which $\langle \sigma |v|\rangle = 1$~pb,
and
the peach-colored and yellow-colored bands around it correspond to the regions 
within which the annihilation cross-section matches this value to within an 
increasing number of powers of ten
(\ie, $0.1\mathrm{~pb} \leq \langle \sigma |v|\rangle \leq 10\mathrm{~pb}$ and 
$0.01\mathrm{~pb} \leq \langle \sigma |v|\rangle \leq 100\mathrm{~pb}$ respectively).
Finally, the blue dashed curve and cyan dot-dashed curve respectively indicate the 
lower limits on $M_{\rm I}$, $M_{\rm II}$, or $M_{\rm III}$ 
from monojet and mono-$W/Z$ searches at the LHC in
the case of a heavy mediator.

Note that we have included the abundance and collider curves within these plots
merely in order to provide guidance when interpreting the impact of the direct-detection curves,
and to indicate regions of specific interest.
In particular, the collider 
and abundance curves do not represent strict bounds in any sense.
For example, within each panel of Fig.~\ref{fig:Abundances},
the region of the $(m_\chi, M_\ast)$ plane {\it below}\/ the $\Omega_\chi \sim  \OmegaDM$   
contour (with $M_\ast$ representing either $M_{\rm I}$, $M_{\rm II}$, or $M_{\rm III}$, as appropriate) 
is actually consistent with observational limits under the assumption that 
some additional contribution makes up the remainder of $\OmegaDM$.  
Conversely, the region {\it above}\/ this contour can also
be consistent with a thermal relic dark-matter candidate if the branching
fraction for dark-matter annihilation into visible-sector particles is less 
than unity due to the presence of additional annihilation channels.
This will also be true if 
an additional source of entropy production dilutes the relic abundance after freeze-out.
Moreover, as discussed in Sect.~\ref{sec:Annihilation},
the abundance-related orange and black curves in Fig.~\ref{fig:Abundances}
do not represent true relic-density limits if $m_\phi < m_\chi$.
Similarly, as discussed in Sect.~\ref{dirdet},
our monojet and mono-$W/Z$ collider curves only represent exclusion bounds under the
assumption that the contact-operator description of our scalar/pseudoscalar interaction
remains valid up to the TeV scale.  When this is not the case, the bounds can 
be far weaker or even effectively disappear.

As we see from Fig.~\ref{fig:Abundances},
the dark-matter abundance does not depend significantly on the
value of $\theta$ in Scenarios~I or III.~
By contrast, in Scenario~II,
our results depend sensitively on $\theta$ due to the enhanced
couplings to third-generation quarks relative to those of the first and
second generations.
Indeed, in Scenario~II the abundance contours have sharp kinks or discontinuities
that are not apparent in Scenarios~I or III.~
This behavior ultimately arises because the couplings to the SM quarks 
in Scenario~II are proportional to their masses,
leading to a dramatic enhancement in the annihilation rate 
when the thresholds for new annihilation channels into heavy quark species are crossed. 
However, this assumes that 
the dark-matter coupling to these heavy-quark species 
is substantial --- a feature that is ultimately $\theta$-dependent.

Overall, examining the plots in Fig.~\ref{fig:Abundances},
we see that there are three main conclusions which may be drawn.
The first and most significant result demonstrated in Fig.~\ref{fig:Abundances}
   is that 
    there are regions of parameter space for which
        a thermal abundance matching $\Omega_{\rm DM}$ is not only consistent with current
     experimental limits on the pseudoscalar operator, but can actually be probed
   by the next generation of direct-detection experiments.
   This does not occur in merely one or two fine-tuned cases, but rather 
   as a fairly generic result for all three scenarios defined 
    in Sect.~\ref{sec:NucleonCouplings} and for most values of $\theta$.

Second, we observe that in some cases, the opposite is true:
     the reach of our direct-detection experiments
     is significantly less than might be expected based
   on the magnitudes of the $\Delta\tilde q^{(N)}$ coefficients.
     This is particularly true for the $\theta=\pi/4$ case 
    of Scenario~I, or the $\theta=0$ case of Scenario~II.~
    Indeed, in such cases, we see that the direct-detection
    experiments cannot even probe that portion of the
       parameter space that would be associated with a thermal relic.
   Moreover, we see from Fig.~\ref{fig:Abundances} that 
    the uncertainties in these cases
    are sufficiently broad that 
   the direct-detection experiments
    may not even have any significant reach at all!
    Ultimately, these effects 
   can easily be understood in relation to Fig.~\ref{fig:gchiNPlot},
   where we have seen that for both of these cases the 
    effective dark-matter/nucleon couplings themselves come extremely close to vanishing.
    (A similar thing would also have happened for
      $\theta\approx \pi/8$ in Scenario~III,
    if such a $\theta$-value were being plotted in Fig.~\ref{fig:Abundances}.)
    As discussed in Sect.~\ref{sec:NucleonCouplings}, 
      these are situations in which the dark matter couples significantly to quarks,
      but not to nucleons.
     In such cases, we conclude that the non-observation of a dark-matter signal
     in COUPP-4 and in future direct-detection experiments need not
    rule out the existence of dark matter which 
    nevertheless still couples to quarks and which could therefore 
     potentially produce a signal at collider experiments.

Finally, conversely, we see that the effects of isospin violation 
      (\ie, variations in the value of $\theta$)
      can have dramatic effects, potentially enhancing the reach of direct-detection
       experiments quite significantly compared with the reach of these experiments
       when nucleon-level isospin symmetry is preserved. 
      For example, in Scenario~I, we note that the reach of PICO-250L is approximately 20 times greater
        (in terms of the values of $M_{\rm I}$ being probed)
      for $\theta=0$ than for $\theta=\pi/4$.


\section{Conclusions}


In this paper, we have studied the sensitivity of direct-detection experiments to dark
matter which couples to quarks through dimension-six effective operators of
the form ${\cal O}_{\chi q}^{\rm (SP)}\sim c_q (\chibar \chi)(\qbar i\gamma^5 q)$, utilizing (for illustrative
purposes) several distinct benchmark choices for the quark couplings $c_q$.
As we discussed, such effective operators give rise to velocity-suppressed spin-dependent dark-matter/nucleon
scattering.  Such operators can also 
give rise to  $\chibar \chi \rightarrow \qbar q$ annihilation from a $p$-wave initial
state, as well as ``mono-anything" signals at the LHC.~

Although it might na\"\i vely be supposed that velocity-dependent spin-dependent scattering 
would produce an unobservably small event rate at direct-detection experiments, we have demonstrated
that this in fact need not be the case.  Indeed, as we have seen, the velocity-suppression factors 
that arise in the pseudoscalar matrix element can be compensated by
extra enhancement factors 
which also emerge in the pseudoscalar case when relating the corresponding
pseudoscalar quark currents 
to effective pseudoscalar nucleon currents.
These latter enhancement factors 
are of size ${\cal O}(10^2-10^3)$
relative to similar factors associated with velocity-{\it independent}\/ spin-dependent scattering
(such as arises through axial-vector interactions).  
As a result, contrary to popular lore,
we see that velocity-suppressed scattering may actually be 
within reach of current and upcoming direct-detection 
experiments.
This then necessitates a sensitivity study of the sort that we have performed.

Specifically, our main conclusions are as follows:
\begin{itemize}
\item  We have shown that there exists a substantial region of $(m_\chi, M_{\rm I, II, III}, \theta)$ 
parameter space in which the couplings of the ${\cal O}_{\chi q}^{\rm (SP)}$  operators are
consistent with a thermal relic density which matches observation.  
Of course, given the model-independent nature of our approach, we have not addressed 
the question of how these operators might ultimately be embedded in a UV-complete model.
Nevertheless, such models can easily be constructed --- for example, the coupling structure of 
Scenario~II can be realized within the context of CP-violating two-Higgs-doublet models~\cite{2HDM}.
\item  A subset of the above parameter space is excluded by current bounds from COUPP-4, and
it is expected that an even larger region of this viable parameter space will be probed by
COUPP-60 and PICO-250L.
\item  While part of the parameter space may be constrained by LHC bounds if 
    the contact-operator approximation remains valid at 
      the TeV scale, there are a wide range of
   models for which spin-dependent scattering is actually the discovery channel.
   As we have seen, this is true because the velocity-suppression effects normally associated with
     pseudoscalar couplings can be overcome through nucleonic effects that emerge
    in relating quark pseudoscalar currents to nucleon pseudoscalar currents.
\item  Conversely, there are special situations (often
    associated with isospin-preserving limits)
    in which these same nucleonic effects render
     direct-detection experiments utterly insensitive
     to non-zero couplings between dark matter and SM quarks.
      In other words, we have seen that dark matter can have a significant,
      non-vanishing coupling to quarks and yet simultaneously have no coupling
       to nucleons!   This opens up the intriguing possibility that
      collider experiments and other indirect-detection experiments
     could potentially see dark-matter signals to which direct-detection
    experiments would be utterly blind.  This may be extremely relevant
    in case of future apparent conflicts between positive signals from 
    collider experiments and negative results from direct-detection experiments. 
\item  Finally, we see that isospin-{\it violation}\/
     generally tends to {\it enhance}\/ dark-matter signals in direct-detection
     experiments relative to the signals which would have been expected 
     if the quark/nucleon couplings were isospin-preserving.
     Moreover, for pseudoscalar couplings,
       this enhancement is not just a factor of two or three 
     (as would be the case for axial-vector interactions),
      but a factor of ten or more.  This then opens up 
      the possibility that direct-detection experiments can 
    be sensitive to such pseudoscalar couplings.

\end{itemize}

A few comments are in order, especially in relation to the last two points above.
In Scenario~I, dark-matter/nucleon couplings are maximally isospin-violating 
when $\theta \approx 0$ or $\theta=\pi/2$.  Interestingly, these cases provide the
greatest sensitivity for direct-detection experiments (such as PICO-250L) which are 
sensitive to spin-dependent scattering.
By contrast, 
detectors which are only sensitive to spin-independent scattering would have no chance
of discovering such events, even if velocity-dependent effects are included.
This is because, as discussed in Sect.~\ref{sec:NucleonCouplings},
all terms originating from
${\cal O}_{\chi q}^{\rm (SP)}$ which contribute to the scattering cross-section ---
and not just those at leading order --- are spin-dependent. 

On the other hand, we have seen that the sensitivity of direct-detection experiments is
especially poor in the isospin-conserving cases 
(such as $\theta \approx \pi/4$ in our Scenario~I)  
for which the couplings to up- and down-quarks are similar.
This poor sensitivity is ultimately the result of a destructive interference
amongst these quark-level couplings, resulting in a small net coupling to both protons and neutrons.
Indeed, in the limit
within Scenario~I for which $m_{u,d} \ll m_s$ and  $\Delta \tilde G^{(N)} \sim 0$, 
we find $g_{\chi p}, g_{\chi n} \rightarrow 0$ identically at $\theta = \pi/4$.  Note that
result is detector-independent:  
one obtains a large suppression in the event rate
 {\it regardless}\/ of whether the detector is sensitive
to spin-dependent scattering from protons or neutrons. 
Moreover, since the dark-matter bilinear is a scalar,
dark-matter annihilation is $p$-wave-suppressed.  
Thus, although the annihilation rate at the time of freeze-out may
have been large enough to ensure the correct relic density, the annihilation
cross-section at the current epoch would be so small as to be unobservable.  
Dark-matter models of this sort
would be difficult to probe via any direct- or indirect-detection
experiments.

In this paper, we considered three different scenarios for the couplings $c_q$ between the
dark matter and SM quarks.  
These correspond to different weightings for the various contributions from the
light quarks to the resulting dark-matter/nucleon couplings and 
their associated dark-matter scattering rates.
In Scenario~I, for example, the dominant contributions 
came from the couplings of the quarks of the first generation, but 
we found that there also exist small contributions from the strange quark and heavier quarks.
Likewise, in Scenario~I we found that $g_{\chi p}, g_{\chi n} \rightarrow 0$ for $\theta = \pi/4$.
In Scenario~III, 
by contrast, the additional contributions from the strange and heavier quarks
are absent.
Moreover, since the $c_q$ coefficients of Scenario~III scale with the masses 
of the quarks, we instead find that 
$g_{\chi p}, g_{\chi n} \rightarrow 0$ for $\tan \theta = m_u / m_d$.
In this connection,
it is perhaps worth emphasizing that it is only for the pseudoscalar interactions 
that there exist values of $\theta$ for which both $g_{\chi p}$ and $g_{\chi n}$ vanish simultaneously.  
As can be seen in Fig.~\ref{fig:gchiNPlot}, this does not happen for any of the analogous couplings
in the axial-vector case.

Finally, Scenario~II is an example of a class of models in which
the largest dark-matter coupling is to the strange quark
or the heavy quarks.
As discussed earlier, this particular example is motivated by minimal flavor-violation.
As evident from Fig.~\ref{fig:Abundances},
the sensitivity of direct-detection experiments to viable
dark-matter models is suppressed for such cases.  It is clear why this occurs.
In Scenario~II, the largest dark-matter couplings are those to the second- and third-generation quarks  ---
indeed, these are ultimately bounded by constraints on the relic density. 
Unfortunately, the contributions from these second- and third-generation quark couplings 
to dark-matter scattering are relatively small 
as a result of a suppression of the corresponding nucleon enhancement factors, 
while the coupling to first-generation quarks is necessarily small by assumption in this scenario.

Depending on the details of the short-distance (ultraviolet) physics model we imagine, 
dark matter which couples to quarks through
an effective operator such as ${\cal O}_{\chi q}^{\rm (SP)}$ 
may also be amenable to ``mono-anything" searches at the LHC.~
In particular, for isospin-conserving variants in which the first-generation quarks dominate
the scattering, LHC searches may be the only viable options for discovery.  Moreover, LHC sensitivity
may be enhanced for flavor structures such as those in Scenario~II which are motivated by minimal 
flavor-violation, 
due to the large contribution to the LHC event rate that arises from the couplings to the heavy quarks.
Ultimately, however, LHC sensitivity depends on the details of the model, and in particular on the flavor
structure of the couplings. 
For the wide class of models 
in which such large LHC event rates do not occur,
spin-dependent direct-detection 
will then be the discovery search channel.


\begin{acknowledgments}


We would like to thank Z.~Chacko, D.~Marfatia, D.~Sanford, 
W.~Shepherd, M.~Sher, X.~Tata, and U.~van~Kolck for useful discussions.
KRD, JK, and BT would also like to thank the Kavli Institute for Theoretical
Physics (KITP) in Santa Barbara, the Galileo Galilei Institute for Theoretical Physics (GGI)
in Florence, Italy,
and the Center for Theoretical Underground Physics and Related Areas (CETUP$^\ast$) in 
South Dakota for their hospitality during various stages of the completion of this work.
JK and BT would also like to acknowledge these institutions for partial support
during these periods.
KRD is supported in part by the Department of Energy under Grants DE-FG02-04ER-41298
and DE-FG02-13ER-41976, and was also supported in part by the National Science Foundation through its
employee IR/D program through August 2013.
The work of JK before June 1, 2013 was supported in part by Department of Energy
grant DE-FG02-04ER41291, while the work of JK after June 15, 2013 is supported in
part by NSF CAREER Award PHY-1250573.
BT is supported in part by DOE Grant DE-FG02-04ER-41291
and in part by the Natural Sciences and Engineering Research Council of Canada.
DY is supported in part by DOE Grants DE-FG02-04ER-41291
and DE-FG02-13ER-42024.
The opinions and conclusions expressed herein are those of the authors, and do not
represent either the Department of Energy or the National Science Foundation.

\end{acknowledgments}





\begin{references}

\bibitem{GoodmanWitten:1985}
  M.~W.~Goodman and E.~Witten,
  Phys.\ Rev.\ D {\bf 31}, 3059 (1985).


\bibitem{JungmanKamionkowskiGriest}
  G.~Jungman, M.~Kamionkowski and K.~Griest,
  Phys.\ Rept.\  {\bf 267}, 195 (1996)
  [arXiv:hep-ph/9506380].

\bibitem{DirectDetReviews}
  D.~Hooper,
  arXiv:0901.4090 [hep-ph];\\
  N.~Weiner,
  ``Dark Matter Theory,'' video of lectures given at TASI 2009,
  {\tt http:// physicslearning2.colorado.edu/tasi/tasi\_2009/ \\
   tasi\_2009.htm}; \\
  J.~L.~Feng,
  Ann.\ Rev.\ Astron.\ Astrophys.\  {\bf 48}, 495 (2010)
  [arXiv:1003.0904 [astro-ph.CO]];\\
  R.~W.~Schnee,
  arXiv:1101.5205 [astro-ph.CO].

\bibitem{LewinSmith}
  J.~D.~Lewin and P.~F.~Smith,
  Astropart.\ Phys.\  {\bf 6}, 87 (1996).


\bibitem{FoxPoppitz:2009}
  P.~J.~Fox and E.~Poppitz,
  Phys.\ Rev.\ D {\bf 79}, 083528 (2009)
  [arXiv:0811.0399 [hep-ph]].

\bibitem{MultiComponentBlock}
  See, {\it e.g.}\/:\\
   D.~Tucker-Smith and N.~Weiner,
   Phys.\ Rev.\ D {\bf 64}, 043502 (2001)
   [hep-ph/0101138];
  Phys.\ Rev.\ D {\bf 72}, 063509 (2005)
  [hep-ph/0402065];\\
  T.~Hur, H.~S.~Lee and S.~Nasri,
  Phys.\ Rev.\ D {\bf 77}, 015008 (2008)
  [arXiv:0710.2653 [hep-ph]]; \\
  J.~L.~Feng and J.~Kumar,
  Phys.\ Rev.\ Lett.\ {\bf 101}, 231301 (2008)
  [arXiv:0803.4196 [hep-ph]]; \\
  H.~S.~Cheon, S.~K.~Kang and C.~S.~Kim,
  Phys.\ Lett.\ B {\bf 675}, 203 (2009)
  [Erratum-ibid.\ B {\bf 698}, 324 (2011)]
  [arXiv:0807.0981 [hep-ph]];\\
   S.~Chang, G.~D.~Kribs, D.~Tucker-Smith and N.~Weiner,
   Phys.\ Rev.\ D {\bf 79}, 043513 (2009)
   [arXiv:0807.2250 [hep-ph]];\\
  K.~M.~Zurek,
  Phys.\ Rev.\ D {\bf 79}, 115002 (2009)
  [arXiv:0811.4429 [hep-ph]]; \\
  B.~Batell, M.~Pospelov and A.~Ritz,
  Phys.\ Rev.\ D {\bf 79}, 115019 (2009)
  [arXiv:0903.3396 [hep-ph]]; \\
  S.~Profumo, K.~Sigurdson and L.~Ubaldi,
  JCAP {\bf 0912}, 016 (2009)
  [arXiv:0907.4374 [hep-ph]]; \\
  F.~Chen, J.~M.~Cline and A.~R.~Frey,
  Phys.\ Rev.\  D {\bf 80}, 083516 (2009)
  [arXiv:0907.4746 [hep-ph]]; \\
  I.~Cholis and N.~Weiner,
  arXiv:0911.4954 [astro-ph.HE]; \\
  X.~Gao, Z.~Kang and T.~Li,
  Eur.\ Phys.\ J.\ C {\bf 69}, 467 (2010)
  [arXiv:1001.3278 [hep-ph]]; \\
  D.~Feldman, Z.~Liu, P.~Nath and G.~Peim,
  Phys.\ Rev.\  D {\bf 81}, 095017 (2010)
  [arXiv:1004.0649 [hep-ph]];\\
  P.~T.~Winslow, K.~Sigurdson and J.~N.~Ng,
  Phys.\ Rev.\  D {\bf 82}, 023512 (2010)
  [arXiv:1005.3013 [hep-ph]];\\
  J.~L.~Feng, M.~Kaplinghat and H.~-B.~Yu,
  Phys.\ Rev.\ D {\bf 82}, 083525 (2010)
  [arXiv:1005.4678 [hep-ph]];\\
  M.~Aoki, M.~Duerr, J.~Kubo and H.~Takano,
  arXiv:1207.3318 [hep-ph];\\
  D.~Chialva, P.~S.~B.~Dev and A.~Mazumdar,
  Phys.\ Rev.\ D {\bf 87}, 063522 (2013)
  [arXiv:1211.0250 [hep-ph]];\\
   S.~Bhattacharya, A.~Drozd, B.~Grzadkowski and J.~Wudka,
   JHEP {\bf 1310}, 158 (2013)
   [arXiv:1309.2986 [hep-ph]].

\bibitem{DDM}
  K.~R.~Dienes and B.~Thomas,
  Phys.\ Rev.\ D {\bf 85}, 083523 (2012)
  [arXiv:1106.4546 [hep-ph]];
  Phys.\ Rev.\ D {\bf 85}, 083524 (2012)
  [arXiv:1107.0721 [hep-ph]];\\
  K.~R.~Dienes, J.~Kumar and B.~Thomas,
  Phys.\ Rev.\ D {\bf 86}, 055016 (2012)
  [arXiv:1208.0336 [hep-ph]].

\bibitem{EllisOliveSavage:2008}
  J.~R.~Ellis, K.~A.~Olive and C.~Savage,
  Phys.\ Rev.\ D {\bf 77}, 065026 (2008)
  [arXiv:0801.3656 [hep-ph]].

\bibitem{Shifman}
  M.~A.~Shifman, A.~I.~Vainshtein and V.~I.~Zakharov,
  Phys.\ Lett.\ B {\bf 78}, 443 (1978).

\bibitem{JaffeManohar}
   R.~L.~Jaffe and A.~Manohar,
   Nucl.\ Phys.\ B {\bf 337}, 509 (1990).

\bibitem{PittelVogel}
   J.~Engel, S.~Pittel and P.~Vogel,
   Int.\ J.\ Mod.\ Phys.\ E {\bf 1}, 1 (1992).

\bibitem{ChackoSpinDep}
  P.~Agrawal, Z.~Chacko, C.~Kilic and R.~K.~Mishra,
  arXiv:1003.1912 [hep-ph].

\bibitem{JiJiMattLianTao}
  J.~Fan, M.~Reece and L.~-T.~Wang,
  JCAP {\bf 1011}, 042 (2010)
  [arXiv:1008.1591 [hep-ph]].

\bibitem{Freytsis:2010ne}
  M.~Freytsis and Z.~Ligeti,
  Phys.\ Rev.\ D {\bf 83}, 115009 (2011)
  [arXiv:1012.5317 [hep-ph]].

\bibitem{Kumar:2013iva}
  J.~Kumar and D.~Marfatia,
  Phys.\ Rev.\ D {\bf 88}, 014035 (2013)
  [arXiv:1305.1611 [hep-ph]].

\bibitem{SMC}
  B.~Adeva {\it et al.}  [Spin Muon (SMC) Collaboration],
  Phys.\ Lett.\ B {\bf 369} (1996) 93;
  Phys.\ Lett.\ B {\bf 420}, 180 (1998)
  [hep-ex/9711008].


\bibitem{HERMES}
  A.~Airapetian {\it et al.}  [HERMES Collaboration],
  Phys.\ Rev.\ D {\bf 71}, 012003 (2005)
  [hep-ex/0407032].

\bibitem{COMPASS}
  M.~Alekseev {\it et al.}  [COMPASS Collaboration],
  Phys.\ Lett.\ B {\bf 660}, 458 (2008)
  [arXiv:0707.4077 [hep-ex]];
  Phys.\ Lett.\ B {\bf 693}, 227 (2010)
  [arXiv:1007.4061 [hep-ex]].


\bibitem{QCDSF}
  G.~S.~Bali {\it et al.}  [QCDSF Collaboration],
  Phys.\ Rev.\ Lett.\  {\bf 108}, 222001 (2012)
  [arXiv:1112.3354 [hep-lat]].

\bibitem{ChengPseudoscalar}
  H.~-Y.~Cheng,
  Phys.\ Lett.\ B {\bf 219}, 347 (1989);\\
   H.~-Y.~Cheng and C.~W.~Chiang,
  JHEP {\bf 1207}, 009 (2012)
  [arXiv:1202.1292 [hep-ph]].

\bibitem{BaiFoxHarnik:2010}
   Y.~Bai, P.~J.~Fox and R.~Harnik,
   JHEP {\bf 1012}, 048 (2010)
   [arXiv:1005.3797 [hep-ph]].

\bibitem{pionpole}
  A.~Kurylov and M.~Kamionkowski,
  Phys.\ Rev.\ D {\bf 69}, 063503 (2004)
  [hep-ph/0307185].

\bibitem{COUPP4Limits}
  E.~Behnke {\it et al.}  [COUPP Collaboration],
  Phys.\ Rev.\ D {\bf 86}, 052001 (2012)
  [arXiv:1204.3094 [astro-ph.CO]].


\bibitem{COUPPdiscussions}
  COUPP Collaboration, private communication.

\bibitem{Brodsky:1988ip}
  S.~J.~Brodsky, J.~R.~Ellis and M.~Karliner,
  Phys.\ Lett.\ B {\bf 206}, 309 (1988).

\bibitem{PDG}
  J.~Beringer {\it et al.}  [Particle Data Group Collaboration],
  Phys.\ Rev.\ D {\bf 86}, 010001 (2012).

\bibitem{student}
  K.~R.~Dienes, J.~Kumar, B.~Thomas, D.~Yaylali {\it et al.}\/, in progress.

\bibitem{monob}
  T.~Lin, E.~W.~Kolb and L.~-T.~Wang,
  Phys.\ Rev.\ D {\bf 88}, 063510 (2013)
  [arXiv:1303.6638 [hep-ph]].

\bibitem{FitzpatrickMathematica}
  N.~Anand, A.~L.~Fitzpatrick and W.~C.~Haxton,
  arXiv:1308.6288 [hep-ph].

\bibitem{SIMPLE}
   M.~Felizardo, T.~A.~Girard, T.~Morlat, A.~C.~Fernandes, A.~R.~Ramos,
    J.~G.~Marques, A.~Kling and J.~Puibasset {\it et al.},
   Phys.\ Rev.\ Lett.\  {\bf 108}, 201302 (2012)
   [arXiv:1106.3014 [astro-ph.CO]].


\bibitem{PICASSO}
  S.~Archambault {\it et al.}  [PICASSO Collaboration],
  Phys.\ Lett.\ B {\bf 711}, 153 (2012)
  [arXiv:1202.1240 [hep-ex]].


\bibitem{COUPPAspenTalk}
  R.~Neilson,  talk given at Aspen 2013,
  {\tt http://indico.cern.ch/getFile.py/access?
  contribId=69{\&}sessionId=3{\&}resId=0{\&}materialId= 
   slides{\&}confId=197862}.

\bibitem{Planck}
   P.~A.~R.~Ade {\it et al.}  [Planck Collaboration],
   arXiv:1303.5076 [astro-ph.CO].

\bibitem{EdsjoGondolo}
  J.~Edsjo and P.~Gondolo,
  Phys.\ Rev.\ D {\bf 56}, 1879 (1997)
  [hep-ph/9704361].

\bibitem{ColliderDMBlock}
   A.~Birkedal, K.~Matchev and M.~Perelstein,
   Phys.\ Rev.\ D {\bf 70}, 077701 (2004)
   [hep-ph/0403004]; \\
   J.~L.~Feng, S.~Su and F.~Takayama,
   Phys.\ Rev.\ Lett.\  {\bf 96}, 151802 (2006)
   [hep-ph/0503117]; \\
   M.~Beltran, D.~Hooper, E.~W.~Kolb, Z.~A.~C.~Krusberg and T.~M.~P.~Tait,
   JHEP {\bf 1009}, 037 (2010)
   [arXiv:1002.4137 [hep-ph]]; \\
   J.~Goodman, M.~Ibe, A.~Rajaraman, W.~Shepherd, T.~M.~P.~Tait and H.~-B.~Yu,
   Phys.\ Lett.\ B {\bf 695}, 185 (2011)
   [arXiv:1005.1286 [hep-ph]]; \\
   P.~J.~Fox, R.~Harnik, J.~Kopp and Y.~Tsai,
   Phys.\ Rev.\ D {\bf 85}, 056011 (2012)
   [arXiv:1109.4398 [hep-ph]];\\
  J.~Goodman and W.~Shepherd,
  arXiv:1111.2359 [hep-ph];\\
  Y.~Bai and T.~M.~P.~Tait,
  Phys.\ Lett.\ B {\bf 723}, 384 (2013)
  [arXiv:1208.4361 [hep-ph]].



\bibitem{ATLASMonojet}
  G.~Aad {\it et al.}  [ATLAS Collaboration],
  JHEP {\bf 1304}, 075 (2013)
  [arXiv:1210.4491 [hep-ex]].

\bibitem{ATLASMonojet8TeV}
  ATLAS Collaboration,
  ATLAS-CONF-2012-147.

\bibitem{CMSMonojet}
  S.~Chatrchyan {\it et al.}  [CMS Collaboration],
  JHEP {\bf 1209}, 094 (2012)
  [arXiv:1206.5663 [hep-ex]].

\bibitem{CMSMonojet8TeV}
  CMS Collaboration,
  CMS-PAS-EXO-12-048.

\bibitem{monoWZ}
  G.~Aad {\it et al.}  [ATLAS Collaboration],
  arXiv:1309.4017 [hep-ex].

\bibitem{MonoCombinedCheung}
  K.~Cheung, P.~-Y.~Tseng, Y.~-L.~S.~Tsai and T.~-C.~Yuan,
  JCAP {\bf 1205}, 001 (2012)
  [arXiv:1201.3402 [hep-ph]].

\bibitem{MonoEverything}
  N.~Zhou, D.~Berge and D.~Whiteson,
  arXiv:1302.3619 [hep-ex].


\bibitem{TimLimits}
  J.~Goodman, M.~Ibe, A.~Rajaraman, W.~Shepherd, T.~M.~P.~Tait and H.~-B.~Yu,
  Phys.\ Rev.\ D {\bf 82}, 116010 (2010)
  [arXiv:1008.1783 [hep-ph]].

\bibitem{MadGraph}
  J.~Alwall, M.~Herquet, F.~Maltoni, O.~Mattelaer and T.~Stelzer,
  JHEP {\bf 1106}, 128 (2011)
  [arXiv:1106.0522 [hep-ph]].

\bibitem{CTEQPDFs}
  J.~Pumplin, D.~R.~Stump, J.~Huston, H.~L.~Lai, P.~M.~Nadolsky and W.~K.~Tung,
  JHEP {\bf 0207}, 012 (2002)
  [hep-ph/0201195].

\bibitem{LowMassMediatorsLHC}
  H.~An, X.~Ji and L.~-T.~Wang,
  JHEP {\bf 1207}, 182 (2012)
  [arXiv:1202.2894 [hep-ph]];\\
  H.~An, R.~Huo and L.~-T.~Wang,
  Phys.\ Dark Univ.\  {\bf 2}, 50 (2013)
  [arXiv:1212.2221 [hep-ph]].

\bibitem{2HDM}
   See, {\it e.g.}\/:\\
    T.~D.~Lee,
   Phys.\ Rev.\ D {\bf 8}, 1226 (1973);\\
   G.~C.~Branco and M.~N.~Rebelo,
   Phys.\ Lett.\ B {\bf 160}, 117 (1985);\\
   S.~Weinberg,
   Phys.\ Rev.\ D {\bf 42}, 860 (1990).\\
   For a review and further references, see\\ 
   G.~C.~Branco, P.~M.~Ferreira, L.~Lavoura, M.~N.~Rebelo, M.~Sher and J.~P.~Silva,
  Phys.\ Rept.\  {\bf 516}, 1 (2012)
  [arXiv:1106.0034 [hep-ph]].





\end{references}
\end{document}